\documentclass[superscriptaddress,nofootinbib,twocolumn,pre]{revtex4-1}

\usepackage{amsmath,graphicx}
\usepackage{color} 
\usepackage{subcaption}

\newcommand{\rr}{{\bf r}}
\newcommand{\kk}{{\bf k}}

\begin{document}

\title{Structural crossover in a model fluid exhibiting two length scales: repercussions for quasicrystal formation}
\author{M.C. Walters}
\affiliation{Department of Mathematical Sciences, Loughborough University, Loughborough, LE11 3TU, UK}
\author{P. Subramanian}
\affiliation{Department of Applied Mathematics, University of Leeds, Leeds LS2 9JT, UK}
\author{A.J. Archer}
\affiliation{Department of Mathematical Sciences, Loughborough University, Loughborough, LE11 3TU, UK}
\author{R. Evans}
\affiliation{H. H. Wills Physics Laboratory, University of Bristol, Bristol, BS8 1TL, UK}

\begin{abstract}
We investigate the liquid state structure of the two-dimensional (2D) model introduced by Barkan et al.\ [Phys.~Rev.~Lett.~{\bf 113}, 098304 (2014)], which exhibits quasicrystalline and other unusual solid phases, focussing on the radial distribution function $g(r)$ and its asymptotic decay $r\to\infty$. For this particular model system, we find that as the density is increased there is a structural crossover from damped oscillatory asymptotic decay with one wavelength to damped oscillatory asymptotic decay with another distinct wavelength. The ratio of these wavelengths is $\approx1.932$. Following the locus in the phase diagram of this structural crossover leads directly to the region where quasicrystals are found. We argue that identifying and following such a crossover line in the phase diagram towards higher densities where the solid phase(s) occur is a good strategy for finding quasicrystals in a wide variety of systems. We also show how the pole analysis of the asymptotic decay of equilibrium fluid correlations is intimately connected with the non-equilibrium growth or decay of small amplitude density fluctuations in a bulk fluid.
\end{abstract}

\maketitle

\section{Introduction}
In this paper we investigate the structure of a one-component model fluid described by a pair-potential that exhibits two distinct length scales. We focus on the particle pair correlations in the uniform fluid, i.e.,\,the radial distribution function $g(r)$, and show that the asymptotic decay, $r\to\infty$ˆž, of this function reflects directly the presence of the two length scales. Specifically, our model system displays the phenomenon of structural crossover whereby the wavelength of the slowest oscillatory decay of $g(r)$ changes discontinuously with state point: there is a sharp line in the phase diagram where the wavelength of the oscillations in $g(r)$ crosses-over from one characteristic length scale to another very different one. For our model, the crossover found in the fluid state provides a clear indicator of the location in the phase diagram where quasicrystals (QC) are expected to form.
 
Structural crossover is a rather general phenomenon. It requires: (i) the presence of two, sufficiently distinct, length scales in the potential function and (ii) that the liquid is sufficiently dense that the pair correlation functions decay in an oscillatory fashion. Liquid mixtures, where the two species are of sufficiently different sizes, are natural candidates for such crossover. The first reported example of structural crossover was for a binary mixture of Gaussian soft-core particles of different sizes with the big-small pair interaction described by a particular mixing rule \cite{archer2001binary}. A few years later, Grodon et al.\ \cite{grodon2004decay, grodon2005homogeneous} reported detailed studies of structural crossover in binary (additive) mixtures of hard-spheres (HS) which prompted experimental investigations, using confocal microscopy, for binary HS-like colloidal mixtures confined to two-dimensions (2D) \cite{baumgartl2007experimental}. The results provided some experimental evidence for cross-over. More recent experiments \cite{statt2016direct}, based on three dimensional (3D) confocal microscopy measurements of the partial radial distribution functions $g_{ij}(r)$ for a binary mixture of PMMA (polymethylmethacrylate) particles suspended in a suitable solvent, point clearly to a sharp structural crossover as the concentration of the mixture is changed. The experimental results \cite{statt2016direct} for the wavelengths of the oscillations are very close to those found in simulation and theory for the corresponding HS mixture. Binary mixtures, with species of different sizes, constitute a clear-cut example where structural cross-over occurs. 

For one component systems the genesis of structural crossover is more subtle. A variety of different physics or chemistry can lead to effective interaction potentials between a pair of colloids, or nanoparticles, that exhibit two significantly different intrinsic length scales. Obvious cases in colloid science are the the effective interactions between charged colloids suspended in a solvent containing non-adsorbing polymers \cite{stradner2004equilibrium, sedgwick2004clusters, campbell2005dynamical, sanchez2005equilibrium}. If the screening length of the solvent is relatively large then there is a repulsion between pairs of colloids at larger separations since the (screened) Coulomb repulsion dominates, but the polymers suspended in the solvent give rise to an additional effective (depletion) attraction when the particles become closer. Potentials of this form are often termed `mermaid' potentials \cite{pini2006freezing, archer2007phase} and can also arise via other physical mechanisms, including 2D fluids of colloidal particles adsorbed at an air-water interface \cite{ghezzi1997formation, sear1999spontaneous}. The competing attraction and repulsion at different ranges can lead to particles exhibiting cluster-formation and microphase-separation, in which the cluster-cluster correlations and the particle-particle correlations give rise to contributions to $g(r)$ having oscillations with two very distinct wavelengths and a distinct peak in the static structure factor $S(k)$ at small but non-zero wavenumber $k$ \cite{archer2007phase, sear1999microphase, imperio2006microphase}. {We should emphasize at this point that the structural crossover in $g(r)$ that we discuss is {\em not} in any way a phase transition; there is no thermodynamic singularity associated with the structural change. In the mermaid systems the observed structural crossover is quite distinct from the microphase-separation which these systems also exhibit. The latter is, at least in three dimensions, a genuine phase transition and not a structural crossover.  Note that there are also examples of structural crossovers in some one-dimensional systems -- see e.g.\ Refs.~\cite{pekalski2013periodic, fu2017assembly}.}

The study in Ref.\ \cite{archer2007model} examined in detail the various different contributions to the decay of $g(r)$ for a model system with a hard core and competing attractive and repulsive Yukawa interactions. It was shown that this model exhibits oscillatory-oscillatory crossover in its supercritical region. There is also a growing literature on simple models of `˜water'€™ that exhibit two distinct length scales, such as the Jagla model pair potential \cite{jagla1999core, jagla2001liquid, xu2011waterlike}, which at larger distances has a soft attraction, with a minimum at a certain value of the inter-particle separation. Additionally, the model has a repulsive ramp potential surrounding a hard core potential at smaller separations, so that if the pressure is high enough, the particles can be closer to one another, defining a second smaller length-scale in the inter-particle correlations.

Our present study is motivated by the recent development of models with two length scales designed to understand the formation of stable quasicrystals (QC) in soft matter \cite{barkan2011stability, archer2013quasicrystalline, barkan2014controlled, archer2015soft}. These built on earlier studies \cite{lifshitz1997theoretical, lifshitz2007soft} based on simple Landau-type local free energy functionals that contain terms involving high-order gradients of the order-parameter. Several work on understanding quasipatterns in Faraday waves \cite{savitz2018multiple, edwards1993parametrically, zhang1997pattern, rucklidge2003convergence, rucklidge2009design, rucklidge2012three, skeldon2015can}. Including high-order gradient terms permits the incorporation of multiple length-scales and significant recent progress has been made in understanding how and why soft matter QC form using such theories \cite{achim2014growth, subramanian2016three, subramanian2017spatially}. This body of work shows clearly that effective pair interactions with two different length scales can stabilise quasiperiodic phases \cite{lifshitz2007soft, barkan2011stability, archer2013quasicrystalline, barkan2014controlled, archer2015soft}. However, it is not known what structural features such potentials might give rise to in the fluid state. Here we study the 2D model originally proposed by Barkan, Engel and Lifshitz (BEL) \cite{barkan2014controlled} and show that the two length scales important for QC formation give rise to structural crossover in the fluid phase and discuss the repercussions. While our results are for the particular BEL model system, we expect our conclusions to apply more generally to other 2D systems that form QC \cite{dotera2014mosaic, pattabhiraman2017effect, pattabhiraman2017periodic}, and when suitably generalized to QC forming systems in 3D, of which there are many.

Our paper is arranged as follows: In Sec.\,\ref{sec2} we describe the BEL model potential. Sec.\,\ref{sec3} describes the integral equation and density functional (DFT) theories we employ to calculate $g(r)$ in the liquid state and how we determine the asymptotic decay of this function using a pole analysis in 2D. In the final part of this Section we present results for structural crossover in the BEL model. In Sec.\,\ref{sec4} we consider the stability of the uniform fluid with respect to density fluctuations, treated in the framework of dynamical DFT, and provide an example of how a dodecagonal QC evolves for a state point where the uniform liquid is unstable. We conclude in Sec.\,\ref{sec5} with a discussion of our results and their implications for understanding QC formation in soft matter systems. 

\section{Model Potential}\label{sec2}

 \begin{figure}[t]
\centering
   \includegraphics[width=0.99\columnwidth]{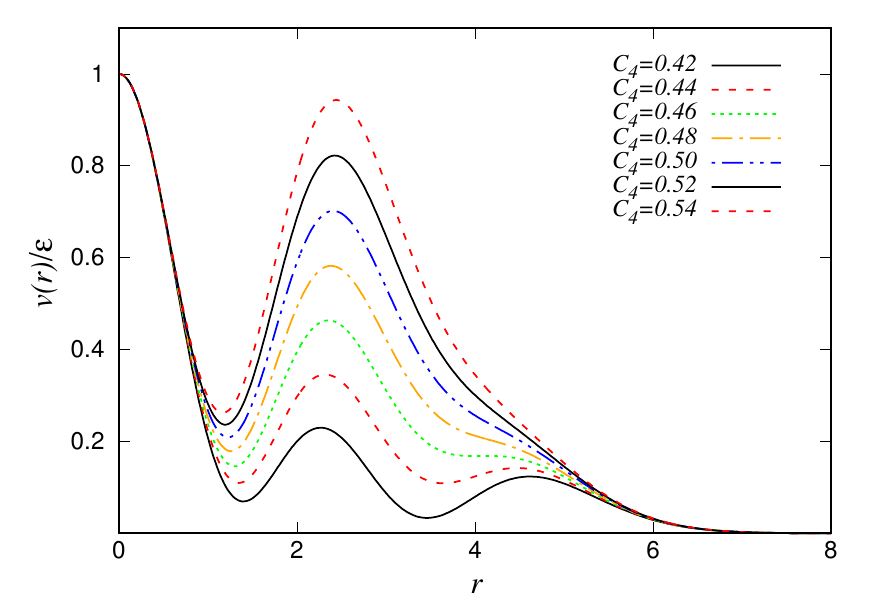}
   \caption{The BEL potential in Eq.\,\eqref{eqn1} where we show the effect of changing only $C_4$ while the other parameters remain constant with values: $\sigma=0.770746$, $C_2=-1.09456$, $C_6=-0.0492739$ and $C_8=0.00183183$. Note that for ${C_4 =C_{4c} =0.439744}$ the two minima in the Fourier transform have identical values, as shown in Fig.\,\ref{fig2}.}
   \label{fig1}
\end{figure}

\begin{figure}[t]
\centering
   \includegraphics[width=0.99\columnwidth]{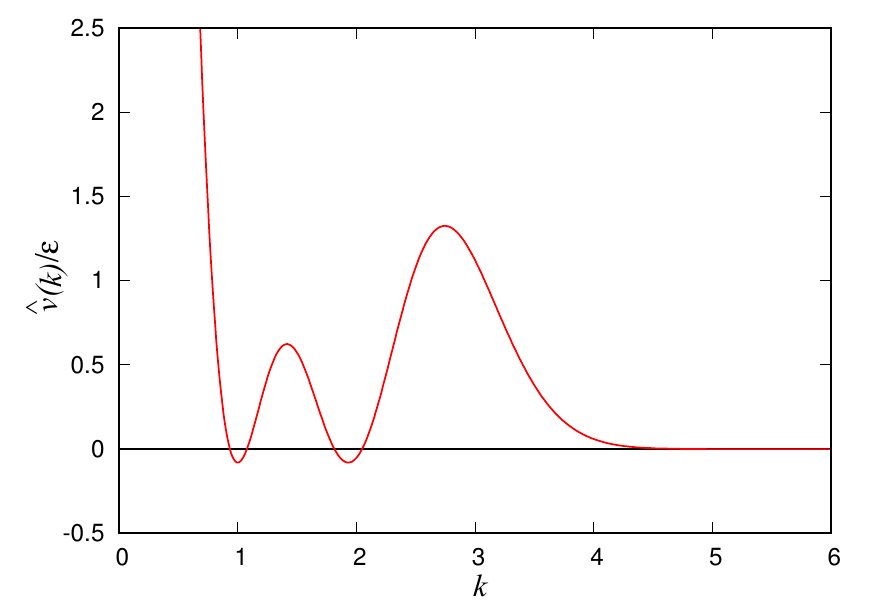}
   \caption{The Fourier transform of the BEL potential with ${C_4 = C_{4c}}$. The two equal minima are at $k_1=1$ and $k_2=1.93185$.}
   \label{fig2}
\end{figure}

We study a 2D system of particles interacting via the model pair potential introduced by Barkan, Engel and Lifshitz  \cite{barkan2014controlled}. The BEL potential combines a Gaussian envelope with a polynomial of order eight:
\begin{equation}
v(r) = \epsilon e^{-\frac{\sigma^2r^2}{2}} \left(1+C_2 r^2 + C_4 r^4+C_6 r^6 + C_8 r^8\right).\label{eqn1}
\end{equation}
The set of coefficients {$C_n$}, for $n = 2, 4, 6$ and $8$, are constants. As in many soft-core models, the energy cost required for one particle to sit directly on top of another is finite and is given by the parameter $\epsilon>0$. The parameter $\sigma$ is the inverse width of the Gaussian and so $\sigma^{-1}$ determines the size of the particles. The potential was constructed first in Fourier space with:
\begin{equation}
\hat{v}(k) = \epsilon e^{-\frac{k^2}{2\sigma^2}}\left( D_0 + D_2 k^2+ D_4 k^4 + D_6 k^6 + D_8 k^8 \right)\label{eqn2}
\end{equation}
where the coefficients {$C_n$} are related directly to the set {$D_n$} \cite{barkan2014controlled}. Barkan et al.\ introduced Eq.~(2) to investigate quasicrystal and other structure formation in 2D. Following earlier work \cite{barkan2011stability}, the authors chose the six coefficients {$D_n$} and $\sigma$ so that the dispersion relation $\omega(k)$ [see Eq.\,\eqref{eqn23}], which determines the growth or decay rate of density modes in the uniform liquid, has two modes which are marginally unstable, one at wavenumber $k_1 = 1$ and second at specified wavenumber $k_2 > 1$. Note that in choosing $k_1=1$ we are setting the larger of the two typical length scales in the system to be $2\pi$. Thus we have effectively nondimensionalised the model, choosing $2\pi$ to be our unit of length.

It is known that if the ratio of the wavenumbers $k_2 /k_1 = 2\cos(\pi/n)$, with integer $n =4,5,6$ or $12$, then stable patterns with $n$ fold symmetry exist in certain models \cite{lifshitz1997theoretical}. Barkan et al.\ \cite{barkan2014controlled} performed a series of molecular dynamics computer simulations, that employed pair potentials \eqref{eqn1} with suitably chosen parameters. Their results exhibited a range of periodic as well as quasiperiodic crystal structures. We focus on the case $n =12$, corresponding to dodecagonal quasicrystalline ordering and shall return to this methodology in Sec.~4.

Fig.\,\ref{fig1} displays the BEL potential in real space for a few values of $C_4$ with the coefficients $C_2$, $C_6$ and $C_8$ held fixed. The values in the caption are those listed by Barkan et al.\ \cite{barkan2014controlled} who showed the choice $C_4 = {C_{4c}}$ produces two identical minima in the Fourier transform at the required ratio of wavenumbers $k_2 /k_1 = 2\cos (\pi/12) = 1.93185$. The BEL potential changes from having two minima at $r\approx 1.5,3.5$ for $C_4=0.42$, to a potential with one minimum at $r\approx 1.2$ for $C_4=0.55$. Clearly the parameter $C_4$ controls the two length scales. Note that $v(r)$ is purely repulsive for these values of the parameters \cite{barkan2014controlled}. In Fig.\,\ref{fig2} we show the 2D Fourier transform for $C_4 ={C_{4c}}$, which exhibits two equal minima at the prescribed wavenumber ratio.

In the next section we investigate how variations in the pair potential, as illustrated in Fig.\,\ref{fig1}, influence structure in the fluid phase. 

\section{Liquid state correlations}\label{sec3}

\subsection{Calculation of $g(r)$}

We focus on the influence of two length scales on pairwise correlations, i.e.\,on the radial distribution function $g(r)$. For soft potentials such as the BEL model the hyper-netted-chain (HNC) approximation \cite{hansen2013theory} is expected to be rather accurate \cite{likos2001effective}. For example, the reliability of the HNC has been established by comparison with simulation for the Gaussian Core Model (GCM) for a wide range of fluid states \cite{likos2001effective}. Moreover, several studies have shown that for the GCM, and closely related Generalized Exponential Models (GEM-n), the simple random phase approximation (RPA) yields results close to those of HNC, especially at high fluid densities \cite{likos2001effective, archer2014solidification}.

\begin{figure*}

\begin{subfigure}{0.48\textwidth}
   \includegraphics[width=1\columnwidth]{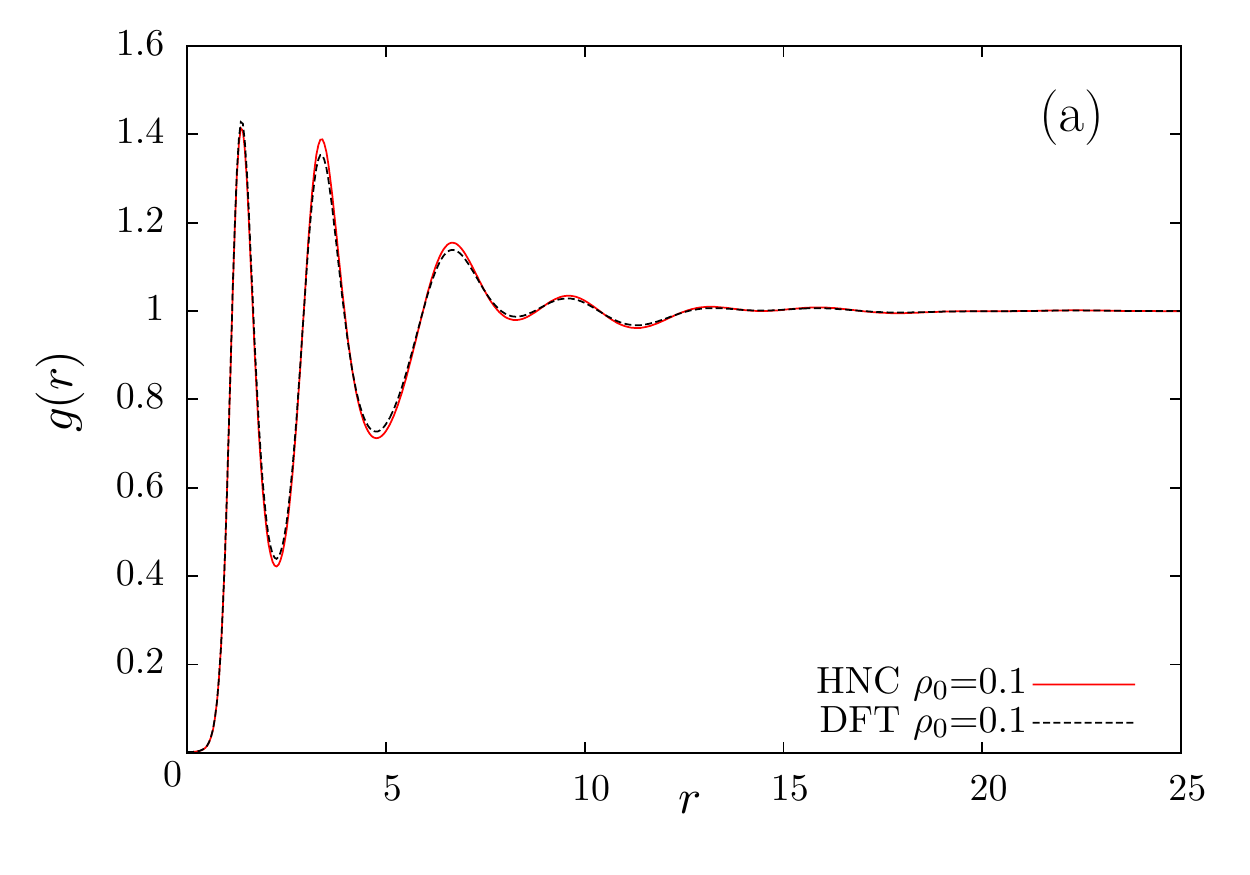}  
\end{subfigure}\hspace{-0.1cm}
\begin{subfigure}{0.48\textwidth}
\vspace{-0.1cm}
 \includegraphics[width=0.99\columnwidth]{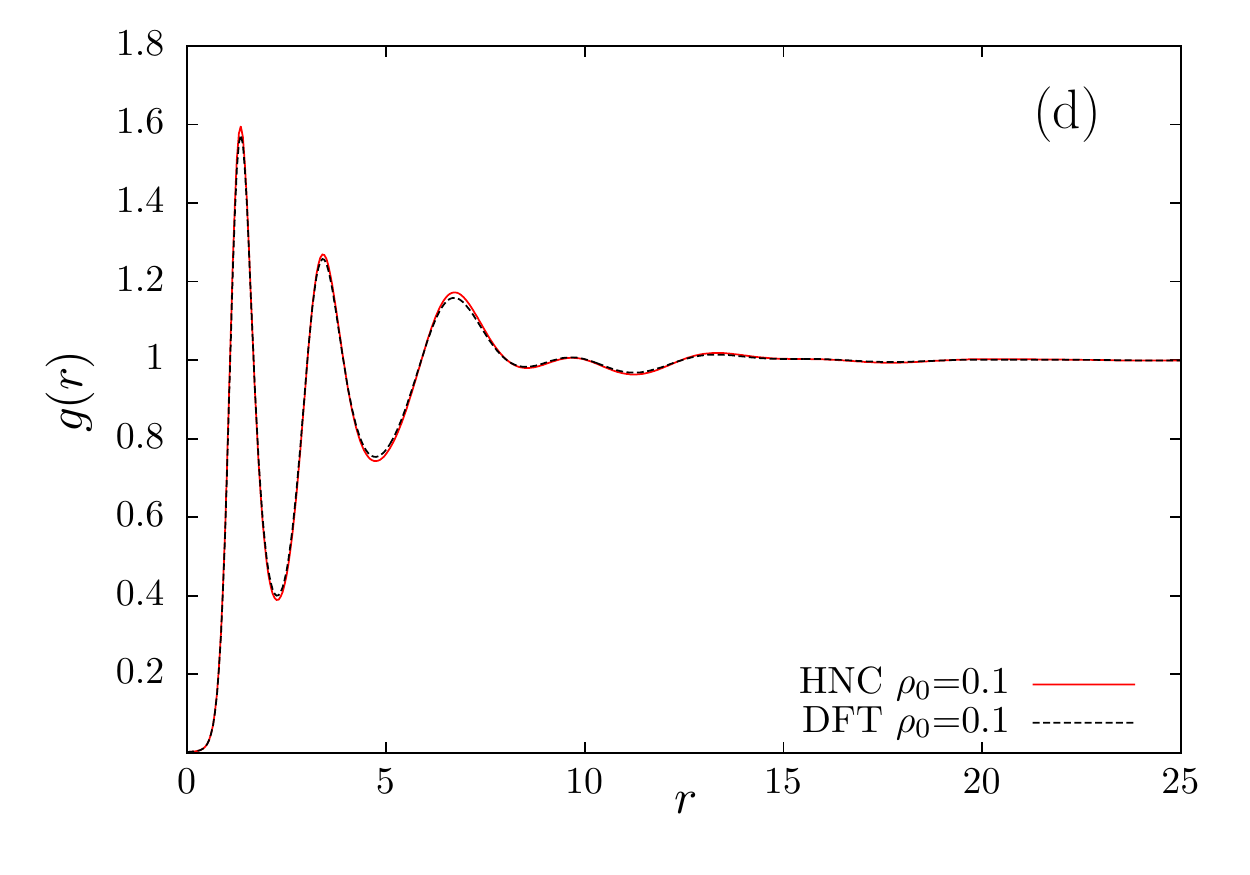}
\end{subfigure}

\medskip
\begin{subfigure}{0.48\textwidth}
\includegraphics[width=1\columnwidth]{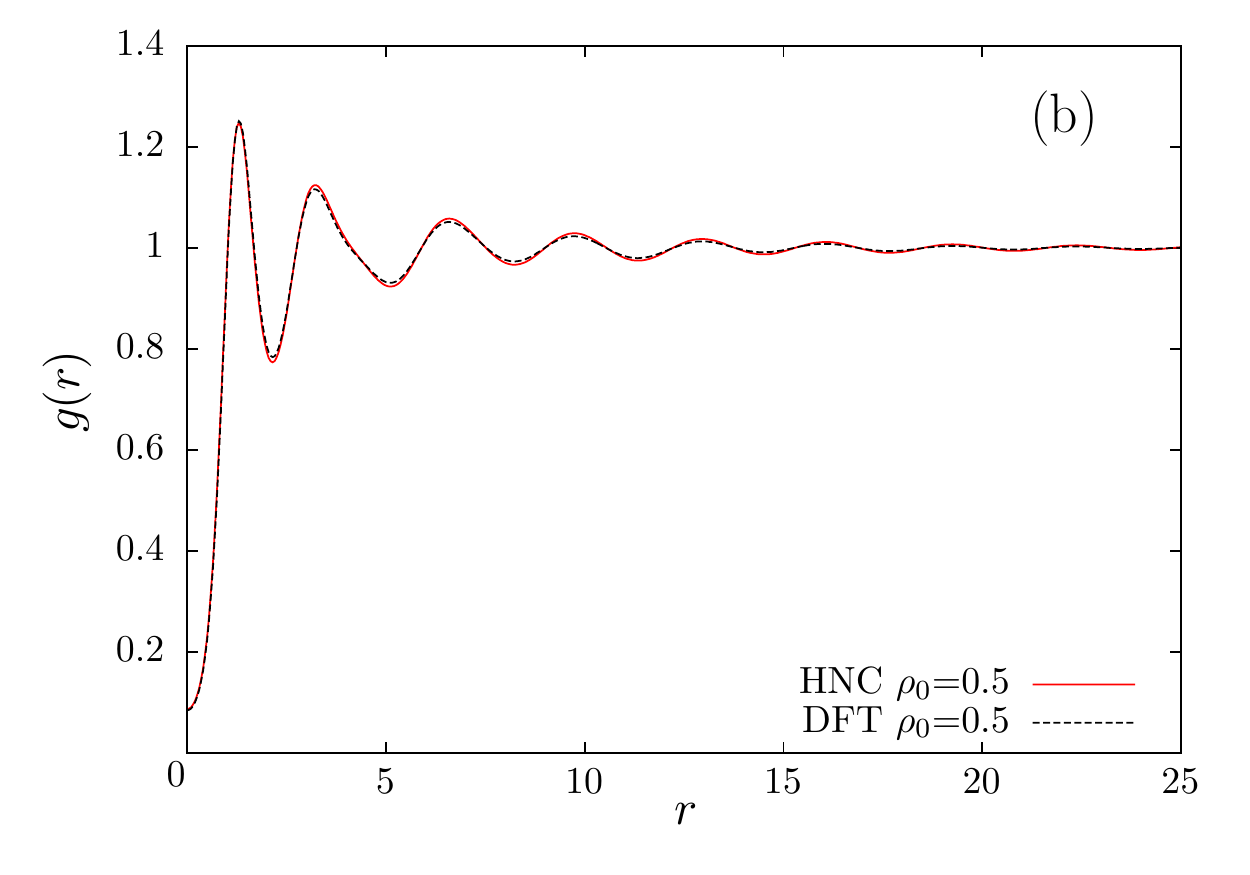} 
\end{subfigure}\hspace*{-0.1cm}
\begin{subfigure}{0.48\textwidth}
\includegraphics[width=1\columnwidth]{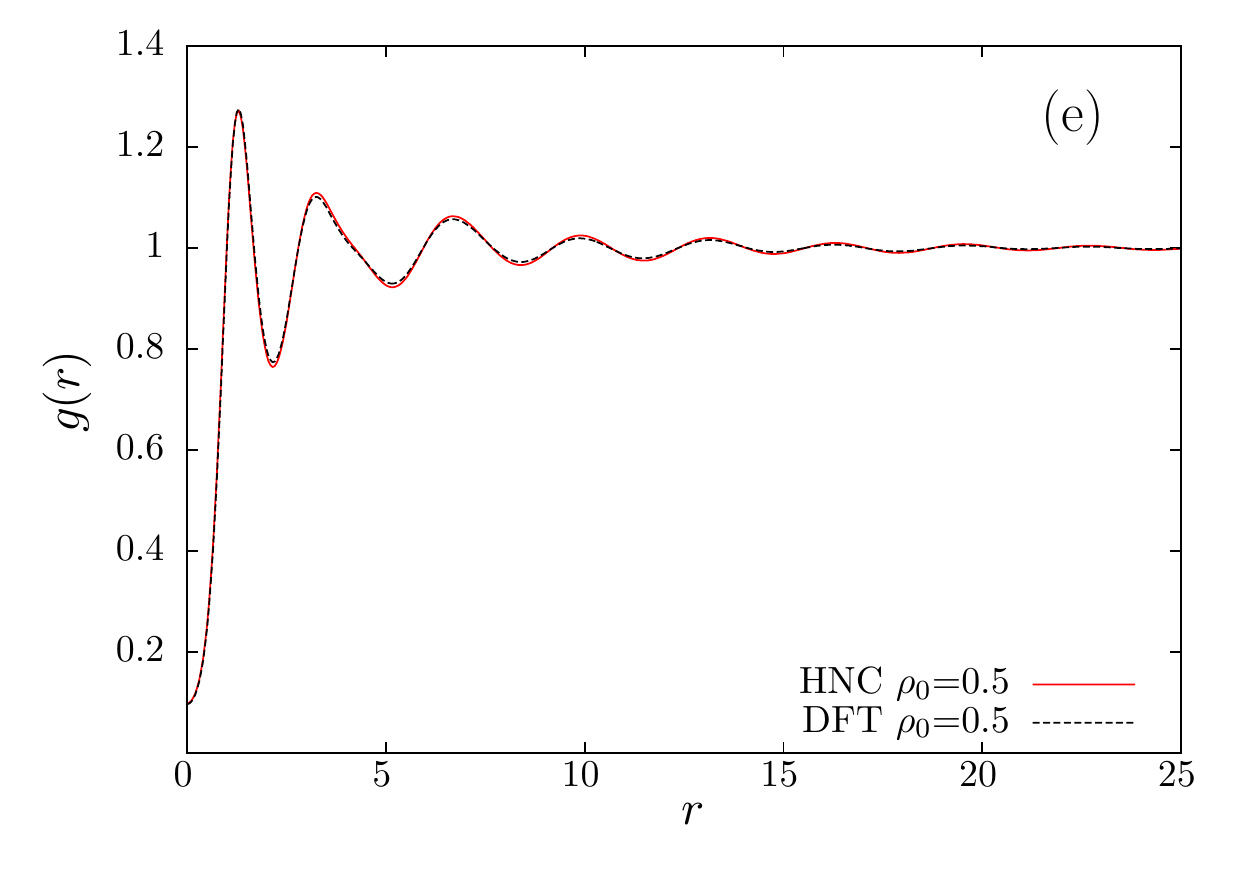} 
\end{subfigure}

\medskip
\begin{subfigure}{0.48\textwidth}
\includegraphics[width=1\columnwidth]{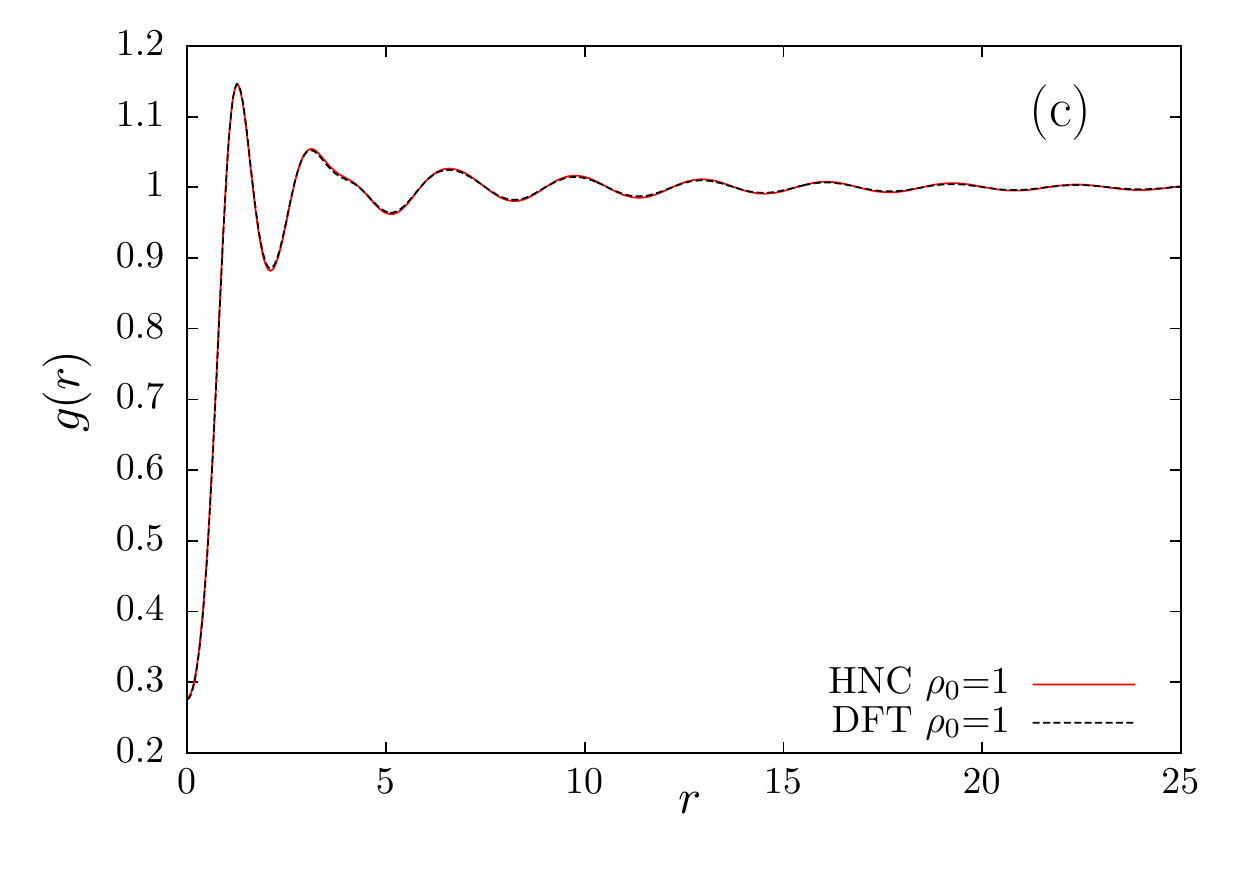} 
\end{subfigure}\hspace*{-0cm}
\begin{subfigure}{0.48\textwidth}
\includegraphics[width=1.02\columnwidth]{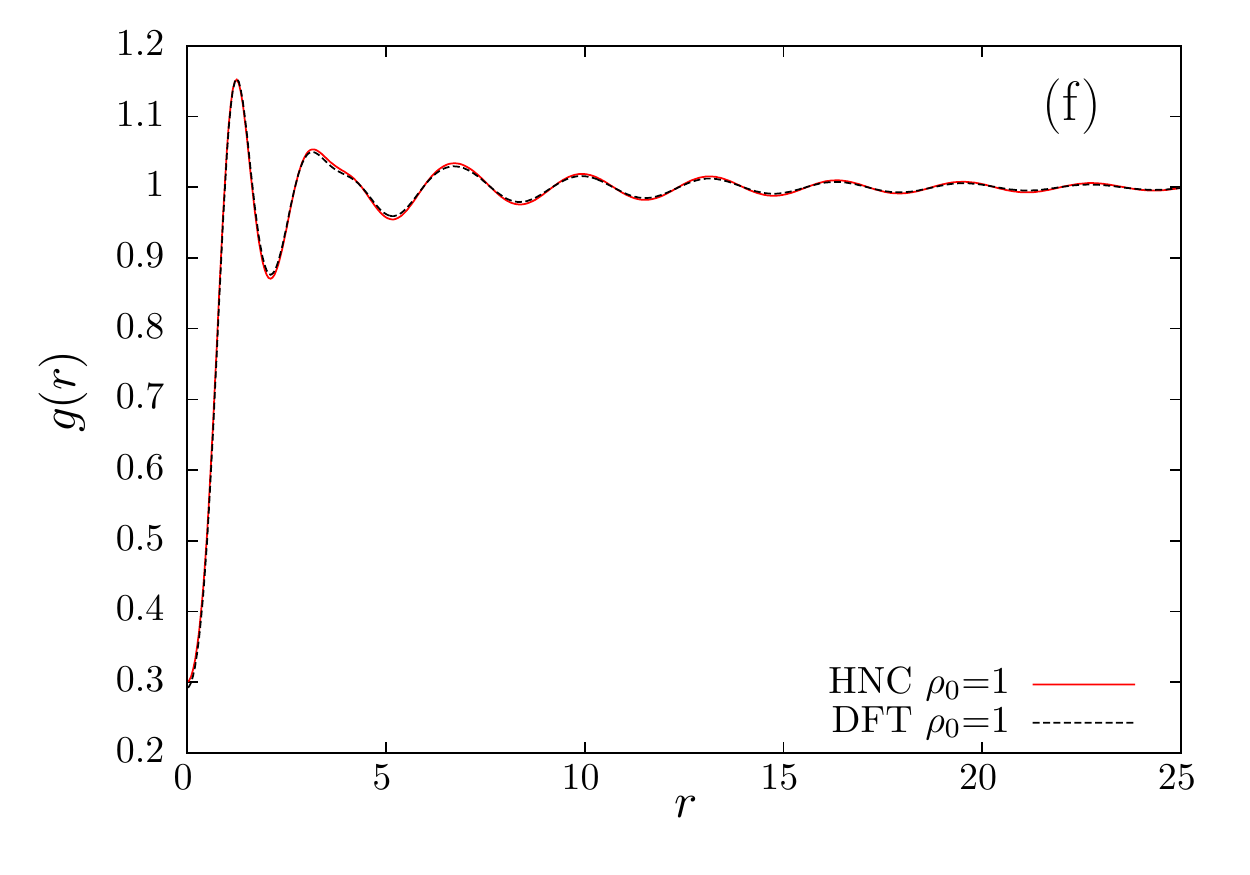}
\end{subfigure}
 
     \caption{{The radial distribution function $g(r)$, (a,b,c) for $C_4=0.42$ and bulk density $\rho_0$ as shown, (d,e,f) for $C_4=0.43$ and bulk density $\rho_0$ as shown.} For all densities there is very good agreement between the HNC and RPA-DFT test particle results. The wavelength of the oscillations at large $r$ changes with density; see text.}
      \label{fig34}
\end{figure*}

For a given (soft) pair potential $v(r)$, the RPA approximates the pair direct correlation function $c(r)$, for all $r$, as
\begin{equation}
c_{RPA}(r) = -\beta v(r)\label{eqn3}
\end{equation}
where $\beta = 1/(k_B T)$, and where $k_B$ is Boltzmann's constant and $T$ is the temperature. In applications it is assumed that the Fourier transform of $v(r)$ exists. The total correlation function $h(r) \equiv g(r)-1$ is then obtained via the (exact) Ornstein-Zernike (OZ) equation, which in Fourier space is~\cite{hansen2013theory}:
\begin{equation}
\hat{h}(k) = \frac{\hat{c}(k)}{1-\rho_0 \hat{c}(k)},\label{eqn4}
\end{equation}
where $\rho_0$ is the bulk density and $\hat{c}(k)$ is the Fourier transform of $c(r)$. The liquid structure factor is defined \cite{hansen2013theory} by $S(k)\equiv 1+ \rho_0 \hat{h}(k)$. It follows that \hbox{$S_{RPA}(k) = [1+\rho_0\beta \hat{v}(k)]^{-1}$} and performing the inverse Fourier Transform yields $g_{RPA} (r)$. This is termed the OZ route. Here we choose to follow another, more accurate, route to $g(r)$, based on classical density functional theory (DFT) \cite{hansen2013theory, evans1979nature, evans1992density} together with the Percus test particle procedure. We build upon the work of Archer et al.\ \cite{archer2014solidification} who investigated the structure of a two-dimensional GEM-4 fluid using the intrinsic Helmholtz free energy functional:
\begin{alignat}{3}
\mathcal{F}[\rho(\rr)]  = & \mathcal{F}_{id}[\rho(\rr)] +  \mathcal{F}_{ex}[\rho(\rr)]\nonumber \\
 = & k_B T\int d\rr \rho(\rr) \left( \ln[\Lambda^2 \rho(\rr)]-1 \right)\nonumber \\
 &+\frac{1}{2} \int d\rr\int d\rr'  \rho(\rr) \rho(\rr') v(|\rr-\rr'|)\label{eqn5}
\end{alignat}
where $\rho(\rr)$ is the one-body density profile with $\rr = (x, y)$. The first term in Eq.~\eqref{eqn5} is the free energy functional of the ideal gas, $\mathcal{F}_{id} [\rho(\rr)]$; $\Lambda$ is the thermal de Broglie wavelength. The second is the excess (over ideal) free energy functional, $\mathcal{F}_{ex} [\rho(\rr)]$, approximated by the standard mean-field form. Taking two functional derivatives of $\mathcal{F}_{ex}$ generates the pair direct correlation function \cite{hansen2013theory, evans1979nature, evans1992density}:
\begin{equation}
c^{(2)}(\rr,\rr') = -\beta\frac{\delta^2 \mathcal{F}_{ex}[\rho]}{\delta \rho(\rr) \delta \rho(\rr')}\label{eqn6}
\end{equation}
and for the approximation \eqref{eqn5} we recover the RPA result Eq.~\eqref{eqn3}. The functional \eqref{eqn5}, and its extension to mixtures, has been used extensively and successfully  in studies of the structure and phase behavior of soft particles \cite{likos2001effective}. Archer et al.\ \cite{archer2014solidification} employed the approximate DFT \eqref{eqn5} in conjunction with the test particle method to calculate $g(r)$. They invoked Percus' \cite{percus1962approximation} result that the one-body density profile {$\rho(\rr)$ around a fixed test particle, exerting on the particles in the fluid an external potential $V_{ext}(\rr)=v(r)$ identical to the pair-interaction potential}, is given by $\rho(\rr) =\rho(r)= \rho_0 g(r)$. By minimizing the grand potential functional, with the approximate intrinsic free energy functional \eqref{eqn5}, one obtains the following integral equation for the density profile and hence $g(r)$ \cite{archer2014solidification}:
\begin{equation}
k_B T \ln\left( \frac{\rho(r)}{\rho_0} \right) + \int d\rr' \rho(\rr') v(|\rr-\rr'|) + v(r) = 0\label{eqn7}
\end{equation}
For the GEM-4 pair potential $v(r) = \epsilon e^{-(r/R)^4}$, where $R$ defines the range, the radial distribution functions obtained from this RPA-DFT test particle route are very close to those from the HNC, even at low temperatures $\beta\epsilon=10$ where one might have expected the approximation to be inaccurate; see Fig.\,1 of \cite{archer2014solidification}.

In Figs.\,\ref{fig34} we display our present results for $g(r)$, with two choices of $C_4$ in the BEL potential \eqref{eqn1}, obtained using this RPA-DFT test particle route alongside those from the HNC approximation. Results are given for fixed $\beta\epsilon=10$ and three values of the (reduced) density $\rho_0$. In all cases there is excellent agreement between the results from the two different approximations. This is remarkable. The BEL potential is much more structured than GEM-4 so one expects much more structured $g(r)$ and it is not obvious that the RPA-DFT should capture the full structure. Recalling that the HNC is generally highly accurate for soft core systems \cite{likos2001effective} these comparisons give us confidence that the RPA-DFT test particle route is a reliable approach and we employ this in the remainder of the paper.

When we compare our results in Fig.\,\ref{fig34} with those in \cite{archer2014solidification} for the GEM-4 potential, which has a single length scale, we glean features associated with two length scales. First, for the higher density states a shoulder develops on the second maximum of $g(r)$ and there is evidence for a `split second peak' at $\rho_0 =1.0$ for $C_4 = 0.42$ and $0.43$. Second, careful observation of the decay of the oscillations in $g(r)$ at large separations shows a significant change in wavelength as the density is increased. For the two higher density states the wavelength is $\approx 0.52\times 2\pi$, for both choices of $C_4$, whereas for $\rho_0 =0.1$ the oscillations are strongly damped but have a much longer wavelength $\approx 2\pi$. We shall account for this observation below. Note that for the two higher densities $g(r=0)$ is greater than zero reflecting the soft-core nature of the pair potential.

\subsection{The asymptotic decay of $h(r)=g(r)-1$: Background}

Important insight into the length scales that determine correlations in the fluid state can be obtained by studying the asymptotic decay, $r\rightarrow\infty$, of the total correlation function $h(r)$. For one-component fluids in 3D the presence of repulsive and attractive portions in the pair potential gives rise to a line in the phase diagram, termed the Fisher-Widom (FW) \cite{fisher1969decay} line after the authors who first pointed to the crossover, whereby the ultimate decay of $h(r)$ crosses-over from monotonic:
\begin{equation}
h(r)\approx \frac{\tilde{A}}{r}e^{-\tilde{\alpha}_0 r},\,\,\,\,\,\,\,\,r\rightarrow\infty~~~~~~~\mathrm{(3D)}\label{eqn8}
\end{equation}
to damped oscillatory
\begin{equation}
h(r)\approx \frac{A}{r}e^{-\alpha_0 r}\cos(\alpha_1r+\theta),\,\,\,\,\,\,\,\,r\rightarrow\infty~~~~~~~\mathrm{(3D)}\label{eqn9}
\end{equation}
FW crossover occurs when decay of type \eqref{eqn8} switches to that of type \eqref{eqn9}, i.e., at a state point where $\tilde{\alpha}_0 = \alpha_0$. Monotonic decay \eqref{eqn8} is found in the neighbourhood of the liquid-gas critical point and in low density gas states whereas exponentially damped oscillatory decay \eqref{eqn9} is associated with high density, liquid or supercritical, states. Such behaviour should be contrasted with the case of one-component HS where the decay is oscillatory for all states. 

The genesis of the two decay types in Eqs.\,\eqref{eqn8} and \eqref{eqn9} emerges from asymptotic analysis of the OZ equation \eqref{eqn4} \cite{evans1993asymptotic, evans1994asymptotic}. Provided the pair potentials are short-ranged the ultimate decay of $h(r)$ is determined by the poles $\alpha$ of $\hat{h}(k)$, i.e., by the solution of $1-\rho_0\hat{c}(\alpha)=0$, with the smallest imaginary part. The poles can be complex: $\alpha=\pm \alpha_1+i\alpha_0$, giving rise to the oscillatory decay in Eq.\,\eqref{eqn9} or purely imaginary $\alpha=i\tilde{\alpha}_0$, giving rise to Eq.\,\eqref{eqn8}. The amplitudes $\mathcal{A}$ and $\tilde{\mathcal{A}}$ are determined by the residues entering the pole analysis \cite{evans1993asymptotic, evans1994asymptotic}. FW crossover occurs at a state point where two distinct, i.e.,  oscillatory and monotonic, branches cross and the imaginary parts are equal. Such a crossover was found in an early DFT study of the square-well model \cite{evans1993asymptotic} and subsequently for a truncated Lennard-Jones potential, using an integral equation approach \cite{de1994decay}. Results for the latter were confirmed in Monte Carlo simulations \cite{dijkstra2000simulation}.

The study by Archer et al.\ \cite{archer2007model}, based on DFT and the Self Consistent Ornstein Zernike Approximation (SCOZA), for a model (mermaid) potential with a double Yukawa potential, attractive at short distances outside the hard core but repulsive at large distances, revealed rich crossover behaviour in the decay of $h(r)$. In the supercritical region of the phase diagram both oscillatory-oscillatory and FW crossover were found. Such complex behaviour arises from the presence of the two different (Yukawa) length scales accompanied by an attractive portion in the pair potential.

As the BEL potential is purely repulsive, intuitively we do not expect to find states exhibiting monotonic decay of $h(r)$. Rather we might expect exponentially damped oscillatory decay for all the states we consider, albeit with the possibility of different wavelengths $2\pi/\alpha_1$. Since our model fluid lives in 2D, we must enquire how the standard 3D pole analysis employed in the studies mentioned above is altered when we consider the lower spatial dimension.

\subsection{A general pole analysis of the asymptotic decay of $h(r)$ in 2D}

We proceed as in 3D by considering the OZ equation (4). The 2D Fourier transform of a function $f(r)$ is given by
\begin{equation}
\hat{f}(k) = 2\pi\int_0^{\infty} dr\, rJ_0(kr)\,f(r)\label{eqn10}
\end{equation}
where $J_0$ is the zeroth Bessel function of the first kind. Similarly, the inverse Fourier transform is
\begin{equation}
f(r) = \frac{1}{2\pi}\int_0^{\infty}\,dk\,kJ_0(kr)\,\hat{f}(k).\label{eqn11}
\end{equation}
It follows that
\begin{equation}
h(r) = \frac{1}{2\pi} \int_0^{\infty} \,dk\,kJ_0(kr)\frac{\hat{c}(k)}{1-\rho_0 \hat{c}(k)}.\label{eqn12}
\end{equation}
We now recall the following asymptotic expansion for the Bessel function:
\begin{eqnarray}
J_0(kr) & = & \sqrt{\frac{2}{\pi kr}}\sin\left(kr+\frac{\pi}{4}\right) + \mathcal{O}\left( \frac{1}{r^{3/2}}\right)\nonumber \\
{~~~~} & = & \sqrt{\frac{1}{\pi kr}} {\cal R}e[(1-i)e^{ikr}] + \mathcal{O}\left( \frac{1}{r^{3/2}}\right)\label{eqn13}
\end{eqnarray}
where ${\cal R}e[z]$ denotes the real part of a complex number $z$. Substituting into Eq.\,\eqref{eqn12} yields
\begin{equation}
h(r) = \frac{1}{2\sqrt{\pi^3r}}\,{\cal R}e[(1-i)\mathcal{I}(r)] + \mathcal{O}\left( \frac{1}{r^{3/2}}\right)\label{eqn14}
\end{equation}
where the integral $\mathcal{I}(r)$ is given by
\begin{equation}
\mathcal{I}(r) \equiv \int_0^{\infty} \, dk\,k^{\frac{1}{2}} e^{ikr}\frac{\hat{c}(k)}{1-\rho_0\hat{c}(k)}\label{eqn15}
\end{equation}
In order to evaluate this integral we convert the integrand into an even function using the substitution $k=\chi^2$. Thus
\begin{equation}
\mathcal{I}(r) = 2\int_0^{\infty}\,d\chi\,\chi^2e^{i\chi^2r}\frac{\hat{c}(\chi^2)}{1-\rho_0\hat{c}(\chi^2)}.\label{eqn16}
\end{equation}

 \begin{figure}[t]
 	\centering
		\includegraphics[width=0.99\columnwidth]{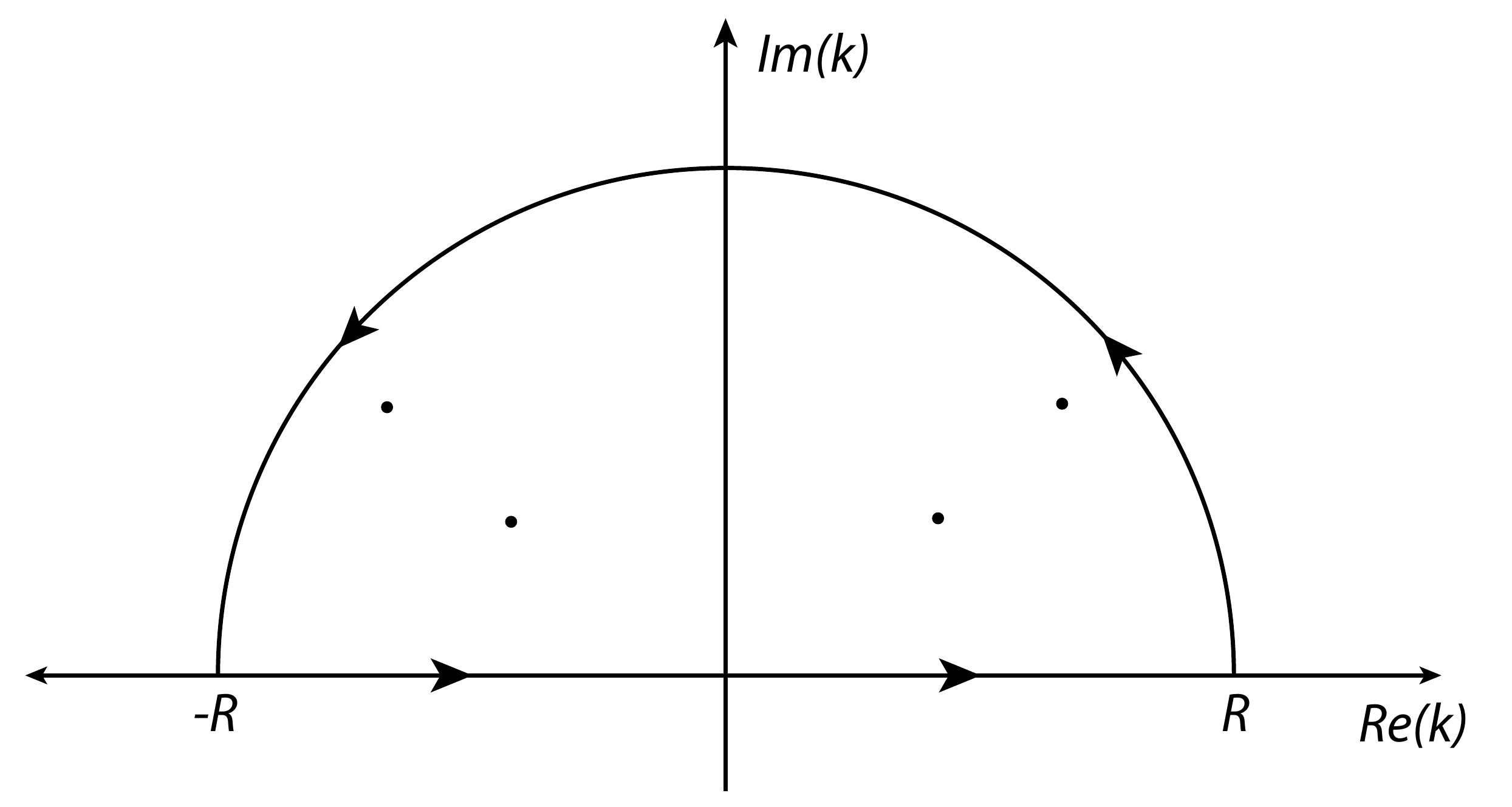}
		\caption{The contour in the complex-$k$ plane used to evaluate the integral in Eq.\,\eqref{eqn16}, in the limit of the circle radius $R\to\infty$. The poles, marked by dots, occur in conjugate complex pairs; see text.}
 	\label{fig5}
\end{figure}

The integral is now of the same form as in 3D and can be evaluated in an analogous manner \cite{evans1993asymptotic, evans1994asymptotic}. This is done by noting that the integrand is even, enabling us to change the limits of the integral $2\int_0^\infty \to \int_{-\infty}^\infty$ and then evaluating the new integral using a closed semi-circular contour in the upper half plane of the complex plane \cite{evans1993asymptotic, evans1994asymptotic}. This is sketched in Fig.\,\ref{fig5}. Using the residue theorem we obtain
\begin{equation}
\mathcal{I}(r) = 2\pi i \sum_n\, R_n\,e^{iq_nr}\label{eqn17}
\end{equation}
where $q_n$ are poles in the upper half-plane, given by the solutions of 
\begin{equation}
1-\rho_0\hat{c}(q_n)=0\label{eqn18}
\end{equation}
and $R_n$ is the residue of $\frac{\chi^2\hat{c}(\chi^2)}{1-\rho_0\hat{c}(\chi^2)}$ at $\chi^2=q_n$.

For short-ranged pair potentials, i.e., those of finite range or those that decay exponentially, or faster, the pair direct correlation function $c(r)$ is also short-ranged, at least for states removed from the bulk critical point. In such cases we expect the poles to be simple. Generally, and by analogy with the 3D fluid, the poles can be pure imaginary $q_n=i\tilde{\alpha}_0$ or come in conjugate complex pairs $q_n=\pm\alpha_1+i\alpha_0$. As in 3D, the slowest decay of $h(r)$ is determined by the pole(s) with the smallest imaginary part. If there is a pure imaginary pole, and $\tilde{\alpha}_0<\alpha_0$, the ultimate decay of the total correlation function takes the form
\begin{equation}
h(r) = \tilde{A}\,\frac{e^{-\tilde{\alpha}_0r}}{\sqrt{r}} + \mathcal{O}\left( \frac{1}{r^{3/2}}\right)~~~~~~~~~~~~\mathrm{(2D)}\label{eqn19}
\end{equation}
where the amplitude $\tilde{A}=\tilde{\alpha}_0{\cal{R}}e[(1-i)/\hat{c}'(i\tilde{\alpha}_0)]/(\sqrt{\pi}\rho_0^2)$ is readily calculated from the residue above. When two conjugate complex poles have $\alpha_0<\tilde{\alpha}_0$ the ultimate decay takes the form
\begin{equation}
h(r) = \frac{Ae^{-\alpha_0r}\,\cos\left( \alpha_1r+\theta \right)}{\sqrt{r}} + \mathcal{O}\left( \frac{1}{r^{3/2}}\right)~~~~~~~\mathrm{(2D)}\label{eqn20}
\end{equation}
where the phase $\theta$ and the amplitude $A$ can be calculated directly from the residues. The calculation mimics that for the 3D case \cite{evans1993asymptotic, evans1994asymptotic}. We see that the asymptotics in 2D follow those in 3D, described by Eqs.\,\eqref{eqn8} and \eqref{eqn9}.  The key difference is the replacement of the factor of $1/r$ in 3D by a factor of $1/\sqrt{r}$ in 2D, reflecting the difference between the Fourier transforms. Given some prescription for the pair direct correlation function $c(r)$ we have a means to determine the asymptotic decay of $h(r)$ in 2D.

\subsection{Poles and structural crossover for the BEL model fluid}

\begin{figure*}
\centering
\includegraphics[width=0.24\textwidth]{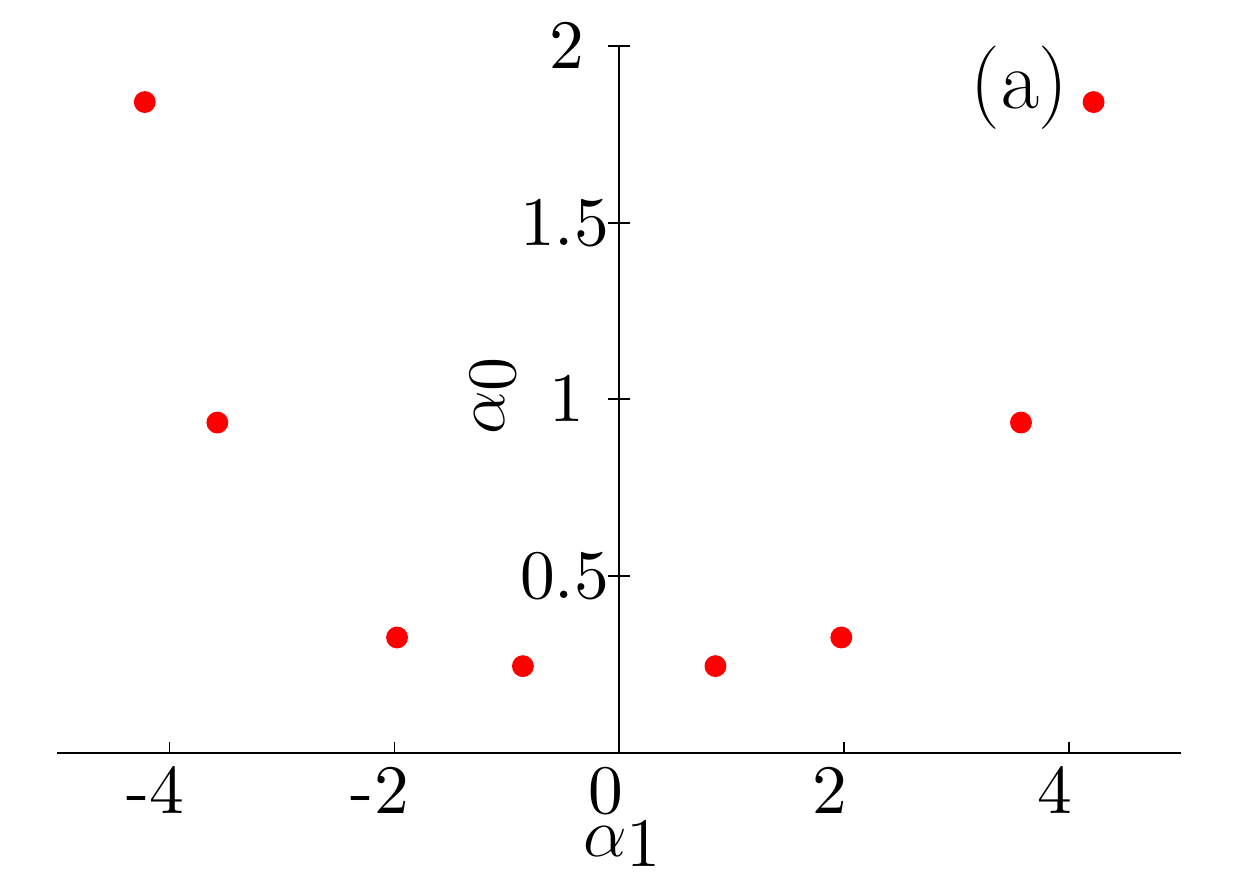}
\includegraphics[width=0.24\textwidth]{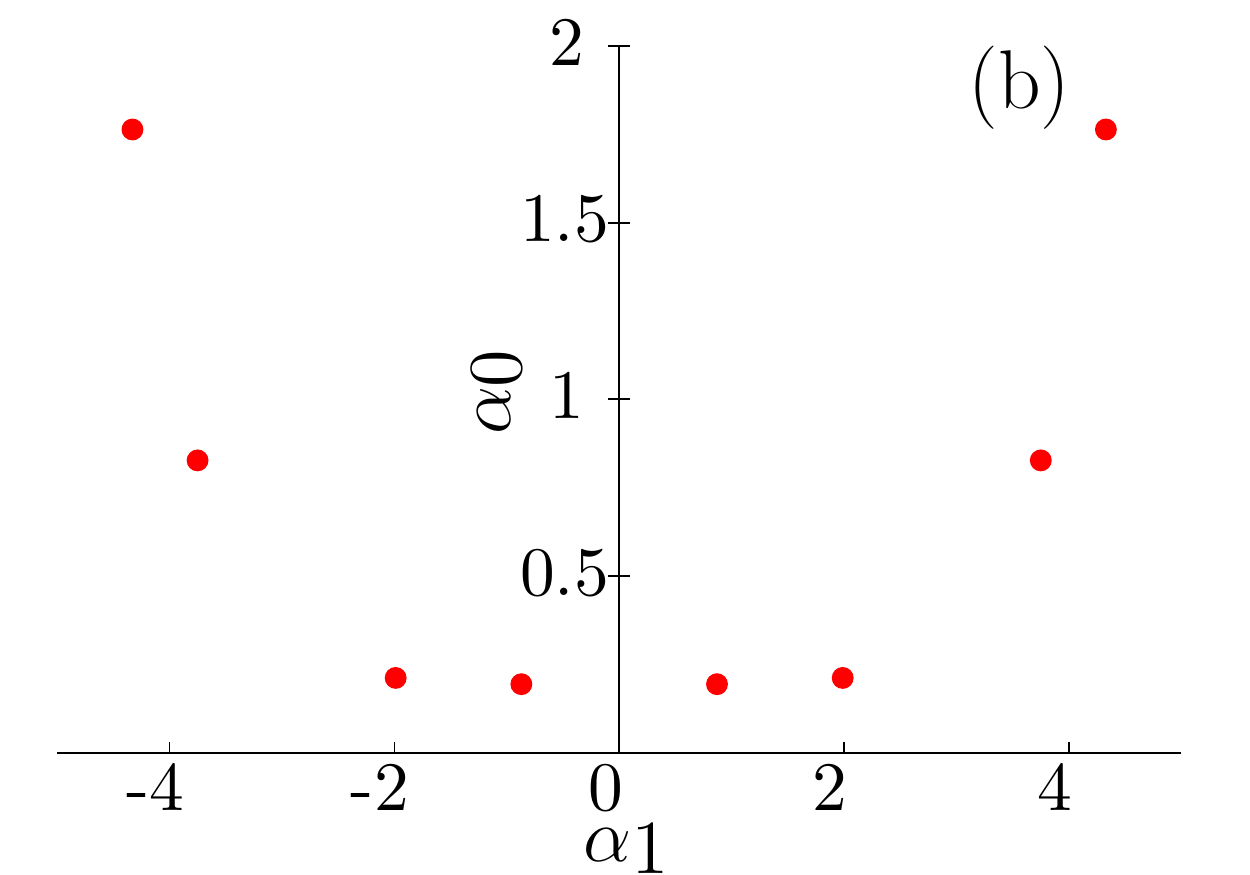}
\includegraphics[width=0.24\textwidth]{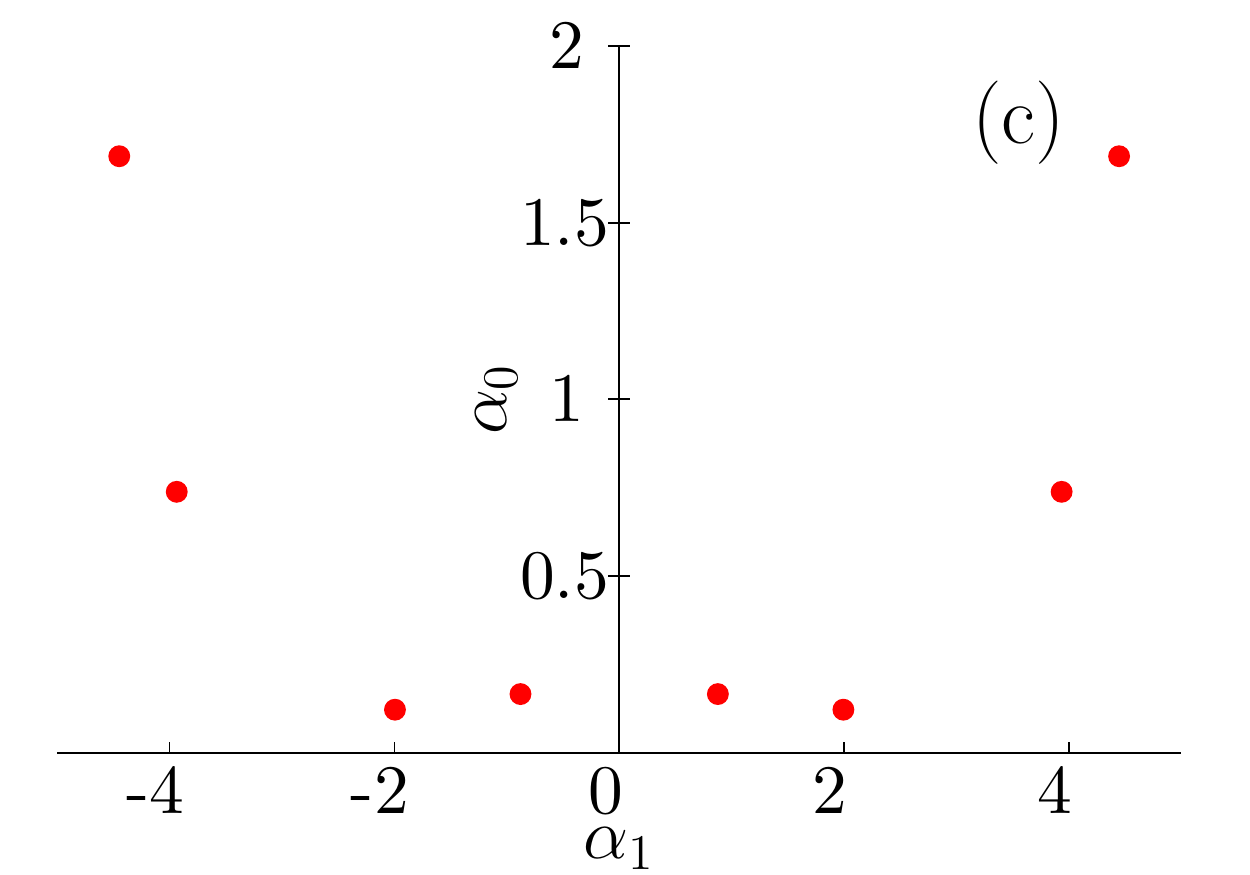}
\includegraphics[width=0.24\textwidth]{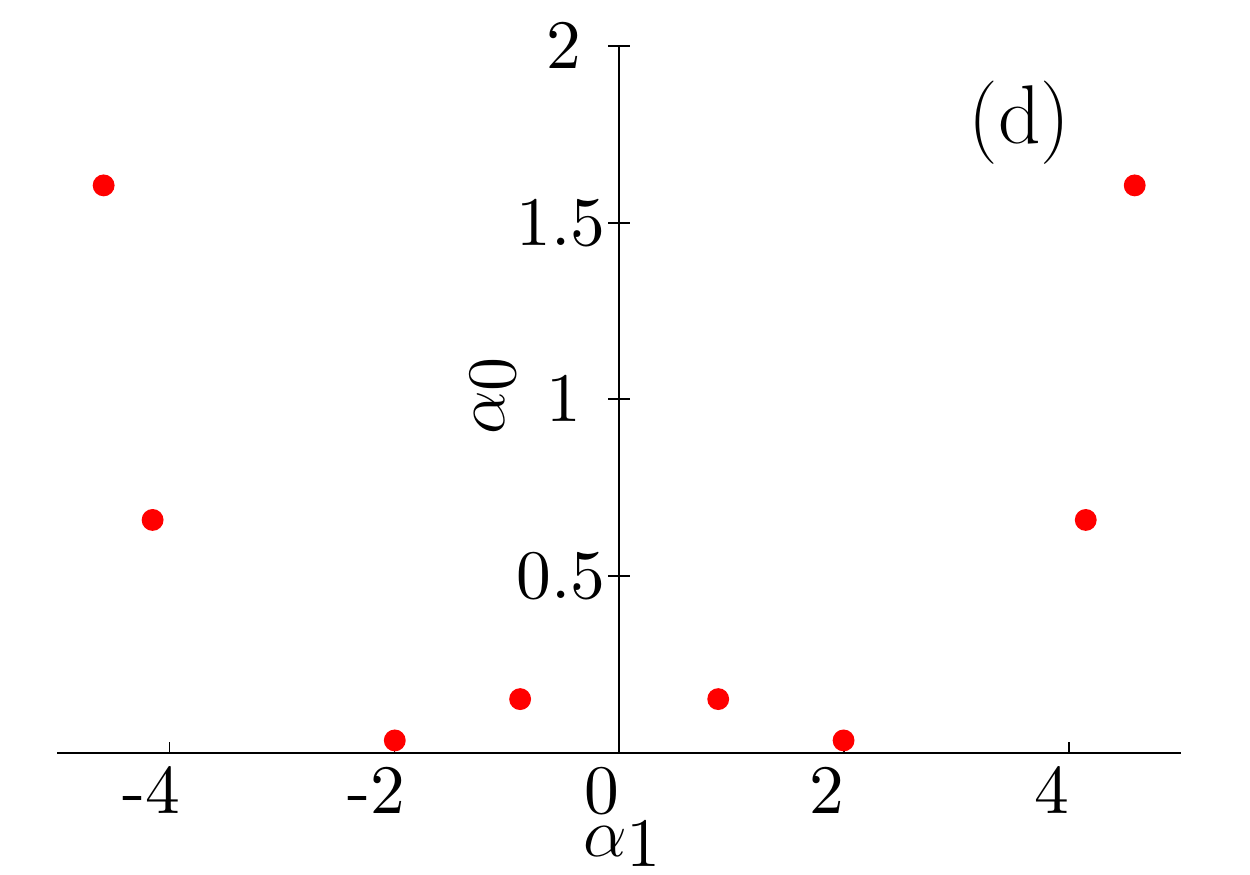}
    
\includegraphics[width=0.24\textwidth]{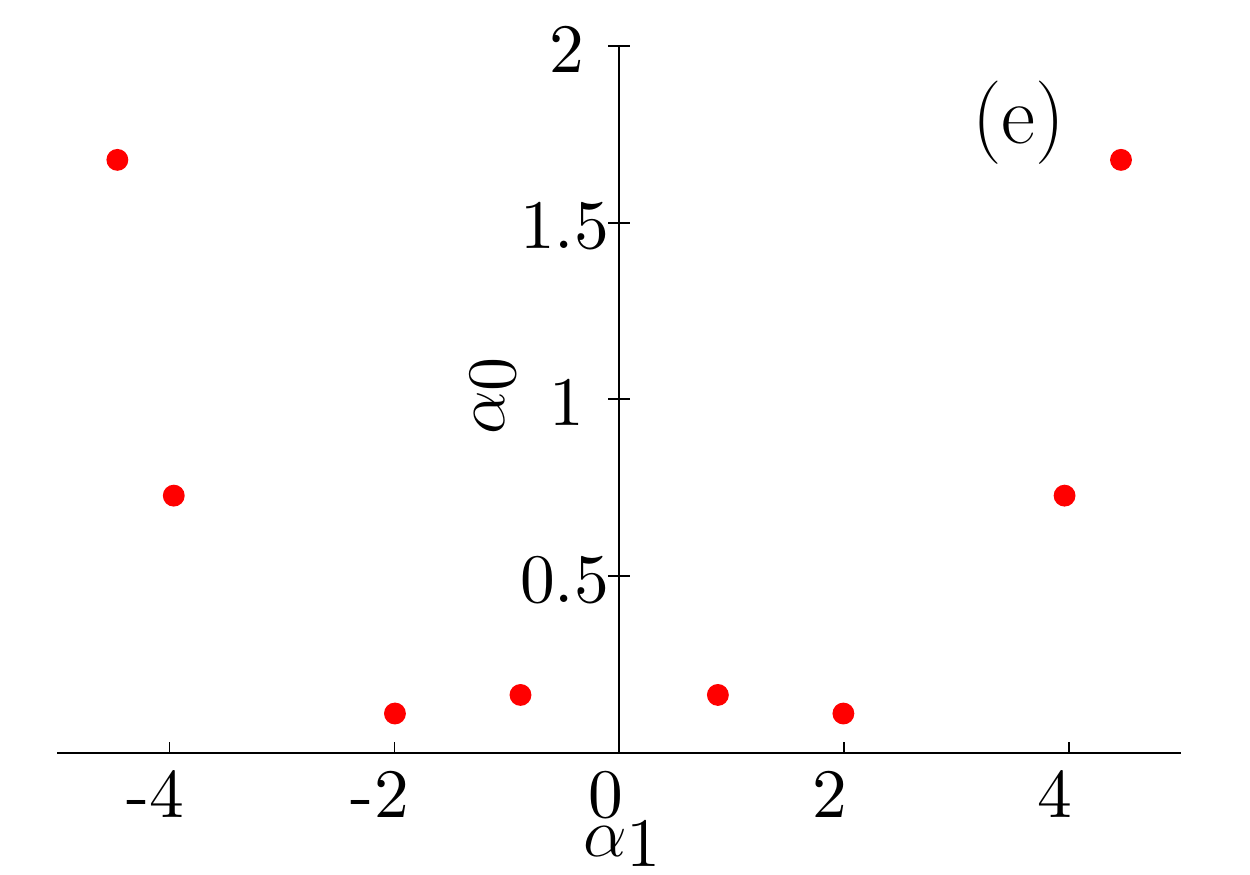}
\includegraphics[width=0.24\textwidth]{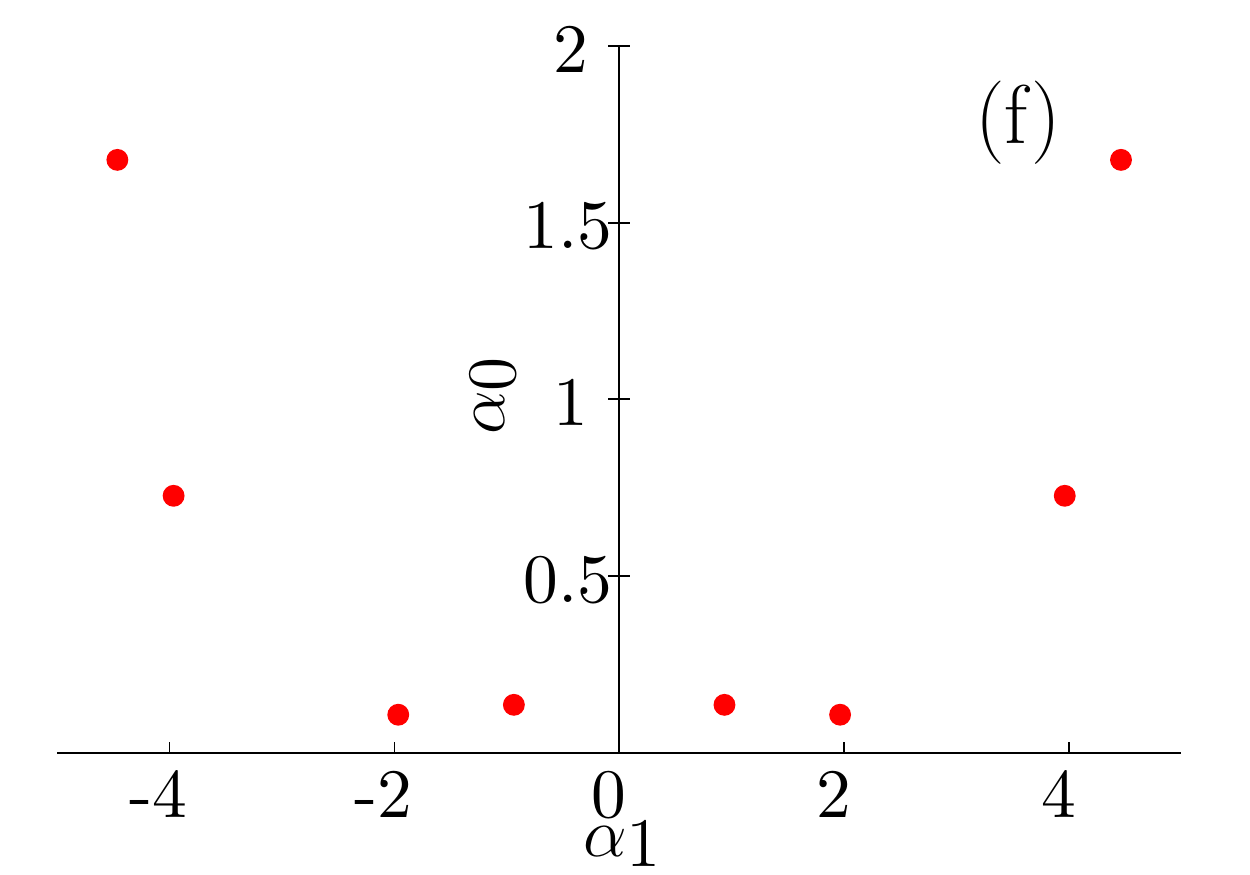}
\includegraphics[width=0.24\textwidth]{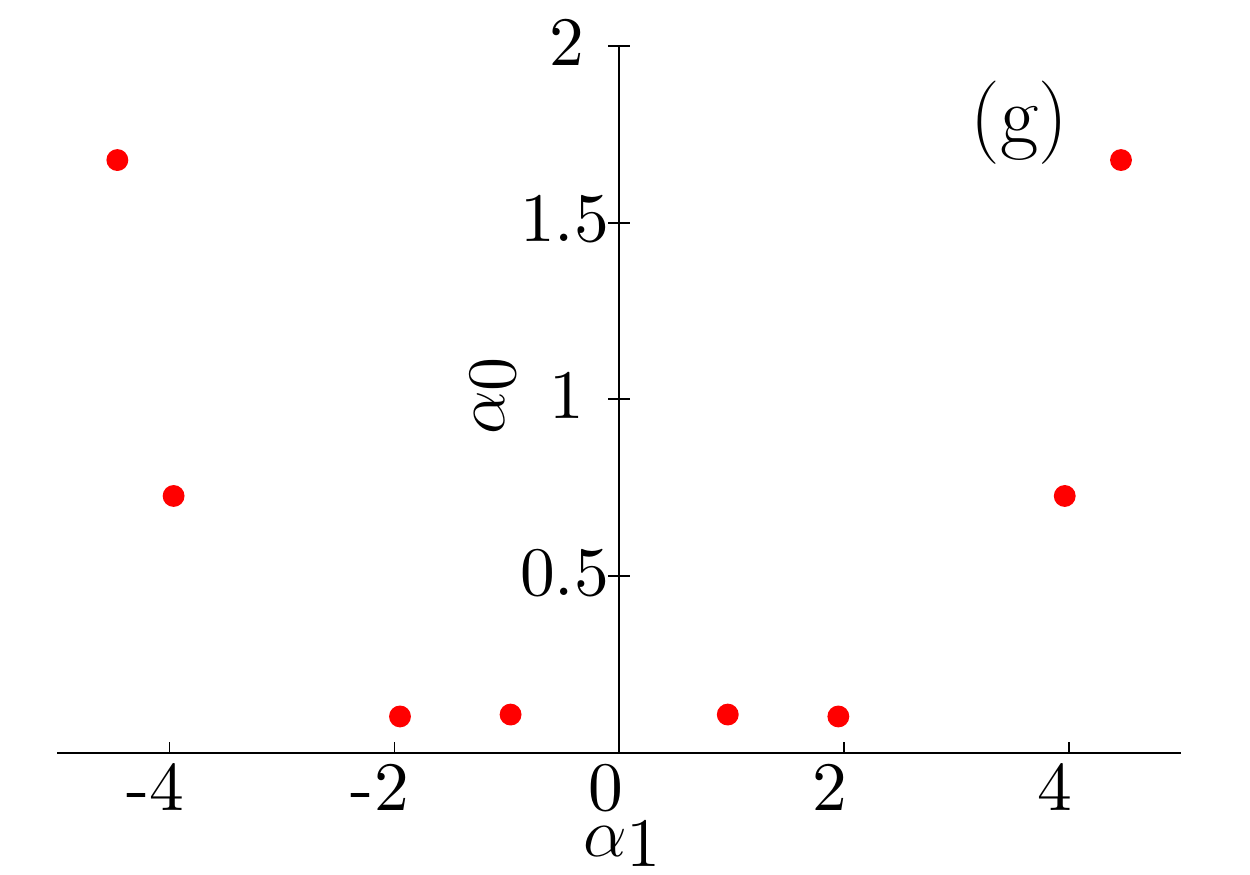}
\includegraphics[width=0.24\textwidth]{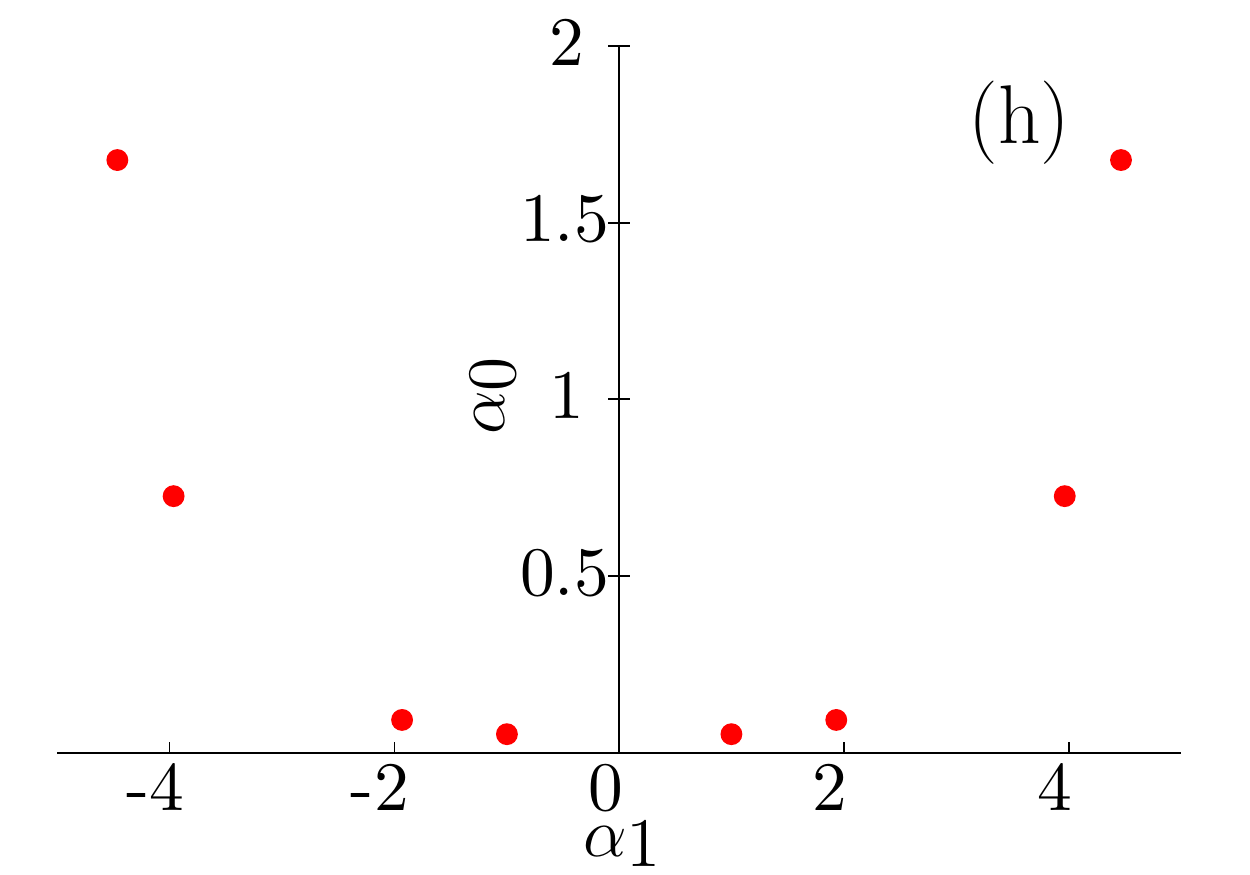}

\includegraphics[width=0.99\columnwidth]{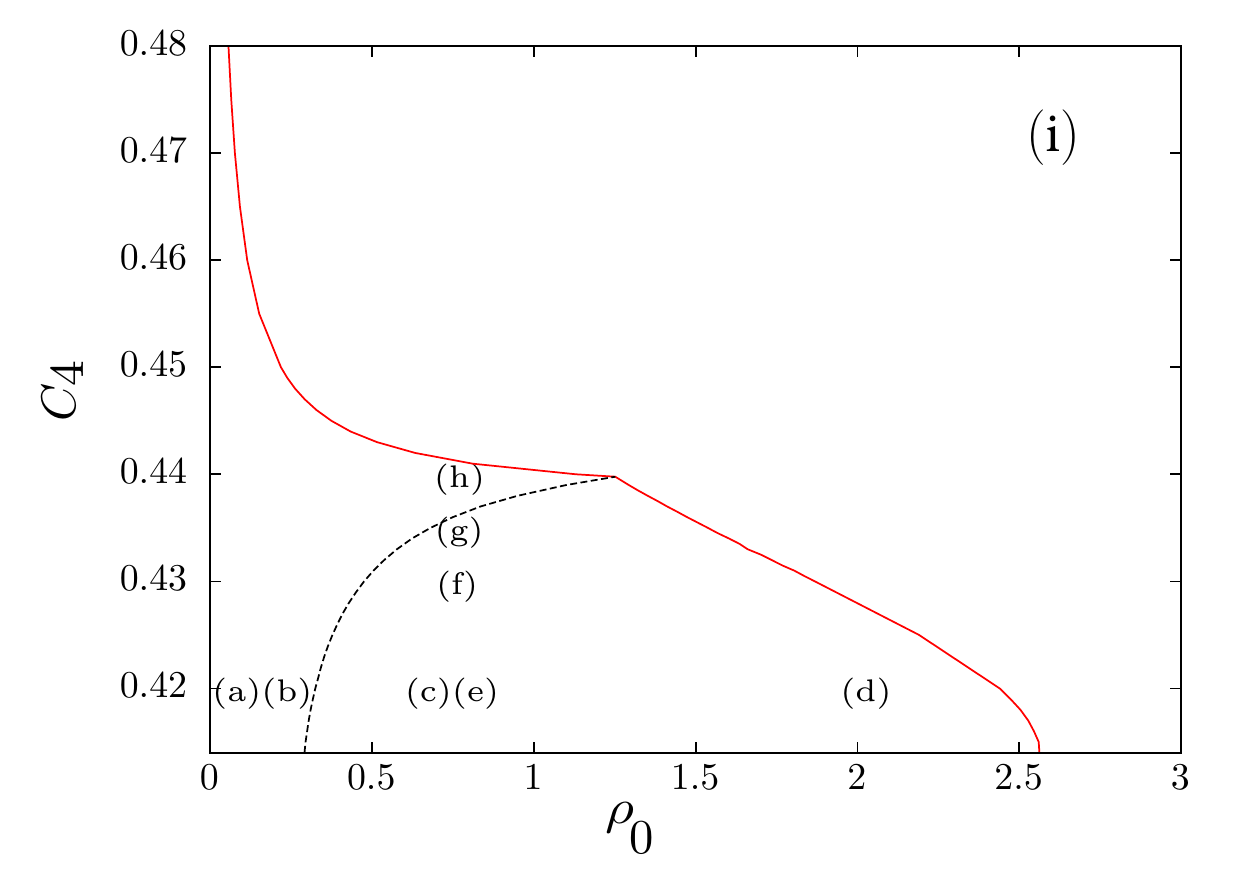}

\caption{Pole structure and structural crossover for the BEL model potential at inverse reduced temperature $\beta\epsilon=10$. The top row displays the lowest lying poles for $C_4 = 0.42$ and reduced densities (a) $\rho_0 =0.1$, (b) $\rho_0 =0.25$, (c) $\rho_0 =0.65$ and (d) $\rho_0 =2.0$. In (a) and (b) the inner poles have the smallest imaginary part whereas in (c) and (d) the outer poles are the lowest lying. The second row displays the poles for $\rho_0 =0.744$ and (e) $C_4=0.42$, (f) $C_4=0.43$, (g) $C_4=0.435$ and (h) $C_4=0.44$. In (e)-(g) the outer poles have the smallest imaginary parts whereas in (h) the inner poles are the lowest lying. The bottom panel {(i)} is the `phase diagram' in the plane which shows the location of the state points (a)-(f) and displays the structural crossover line (black dashed line) where the inner and outer poles have identical imaginary parts. To the left of this line the slowest oscillatory decay of $h(r)$ has a longer wavelength than to the right. The two red solid curves denote the onset of instability of the uniform fluid; the low-density branch corresponds to the mode with $k\approx1$ first becoming linearly unstable while the high-density branch corresponds to the $k\approx1.93$ mode. The structural crossover line runs into the point where the two branches meet at $C_4={C_{4c}}$; see text.}
   \label{fig6}
\end{figure*}

We calculate the poles, as determined by Eq.\,\eqref{eqn18}, using the direct correlation function given by the simple RPA, Eq.\,\eqref{eqn3}. It is important to recognize that the same inverse decay length {and wavelength} characterizing the decay of $h(r)$ arises in the test particle procedure described in Eq.\,\eqref{eqn7}. The equivalence between the test particle and the OZ routes for the length scales of the asymptotic decay is general and is based upon linear response arguments; see \cite{EvansCarvalho96}. Note, however, that the amplitudes and phases will differ between the two routes. We do not attempt to calculate these quantities in this paper.

\begin{figure}[t]
\centering
   \includegraphics[width=0.99\columnwidth]{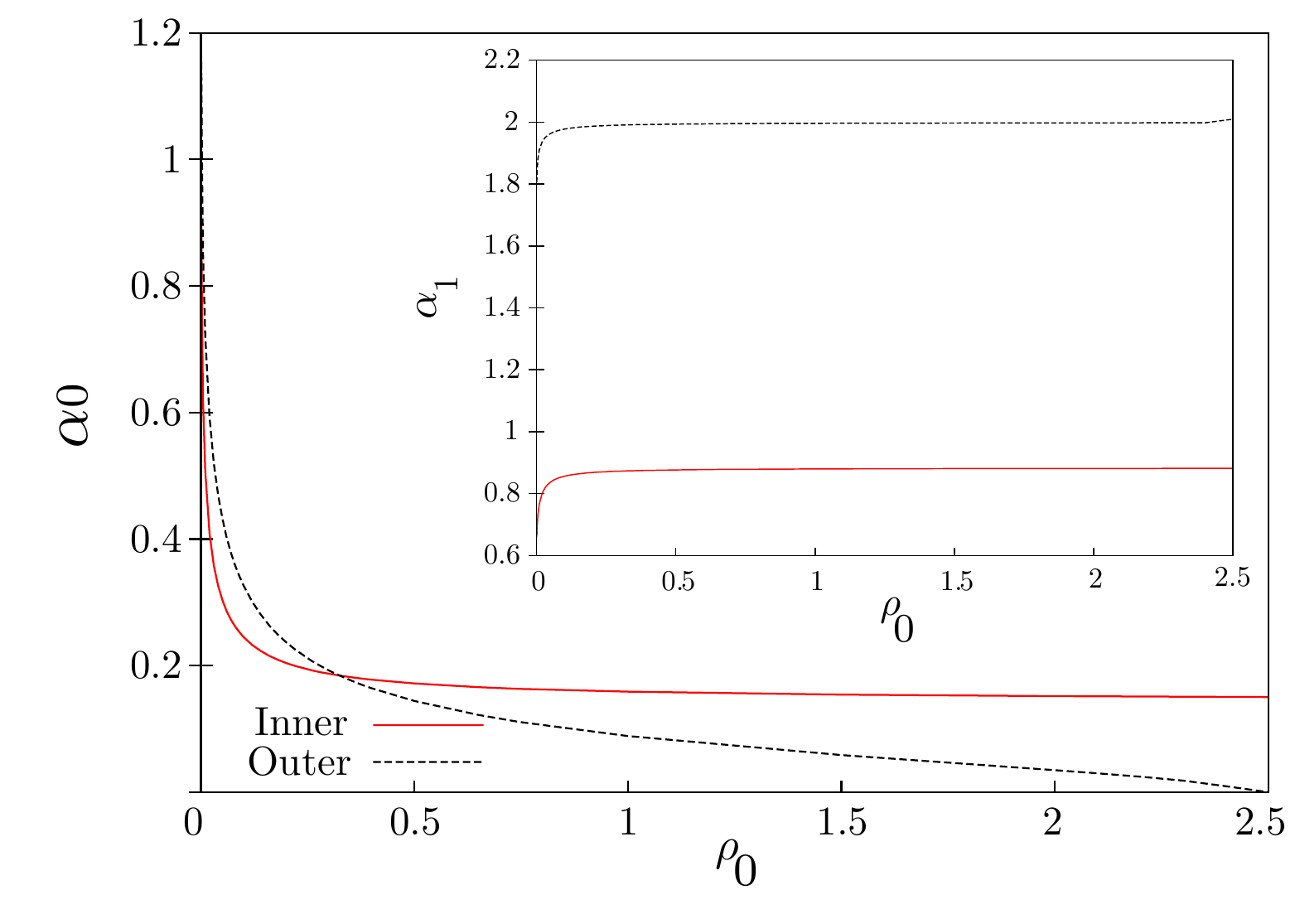}
     \caption{Variation of $\alpha_0$, the imaginary part of the inner poles (red line) and outer poles (green dashed line), with reduced density $\rho_0$ for $C_4 = 0.42$ and $\beta\epsilon=10$. The inset shows the corresponding plot for $\alpha_1$, the real part.}
      \label{fig7}
\end{figure}

For the BEL potential \eqref{eqn1} we find only complex poles, as expected for a purely repulsive, short-ranged pair potential. Examples of the low-lying poles, i.e., with those the smallest values of $\alpha_0$, are shown in Fig.\,\ref{fig6} for various state points. The top row shows the pole structure for fixed $C_4 =0.42$ at four different densities and fixed $\beta \epsilon=10$ while the second row shows the poles for fixed reduced density $\rho_0 =0.774$ and several values of $C_4$ for the same reduced temperature. We define the inner poles as the conjugate pair closest to the $\alpha_1=0$ axis and the outer as the next closest pair. In the top row (a) to (d), we see that the inner pole has the smaller imaginary part at low densities while the outer pole has the smaller imaginary part at high densities. In the second row (e) to (h), we find that the outer pole has the smaller imaginary part for small $C_4$ while the inner acquires the smaller imaginary part at large $C_4$. A state point  at which the imaginary parts of the inner and outer poles are identical is a point of structural crossover. The locus of these points can be plotted in a `phase diagram', as shown in the bottom panel for fixed $\beta\epsilon=10$. The black dashed line is the structural crossover line calculated for the BEL model. This line is the locus of state points where the slowest decay of $h(r)$ switches discontinuously from damped oscillatory with a long wavelength $2\pi/\alpha_1$ (the inner pole is lowest lying) to decay with a shorter wavelength (the outer pole is lowest lying) on increasing the fluid density. The crossover is illustrated in Fig.\,\ref{fig7} for fixed $C_4 =0.42$. For densities $\rho_0<0.35$, the crossover value, the inner pole has the smaller imaginary part and the wavelength of the slowest decaying oscillations is $\approx2\pi$, whereas for larger densities the outer pole is the lowest lying and the wavelength is $\approx0.52\times2\pi$. This result accounts for the different wavelengths of decay observed in the RPA-DFT and HNC results for $g(r)$ in Fig.\,\ref{fig34}. {This observation should aid in identifying new systems that exhibit QC formation. By finding such a crossover line with the correct length scale ratios in the {\em liquid} state portion of the the phase diagram and following it towards where the solid phases exist, one is heading to the portion of the phase diagram where QCs are most likely to occur}.

 \begin{figure}
\centering
   \includegraphics[width=0.99\columnwidth]{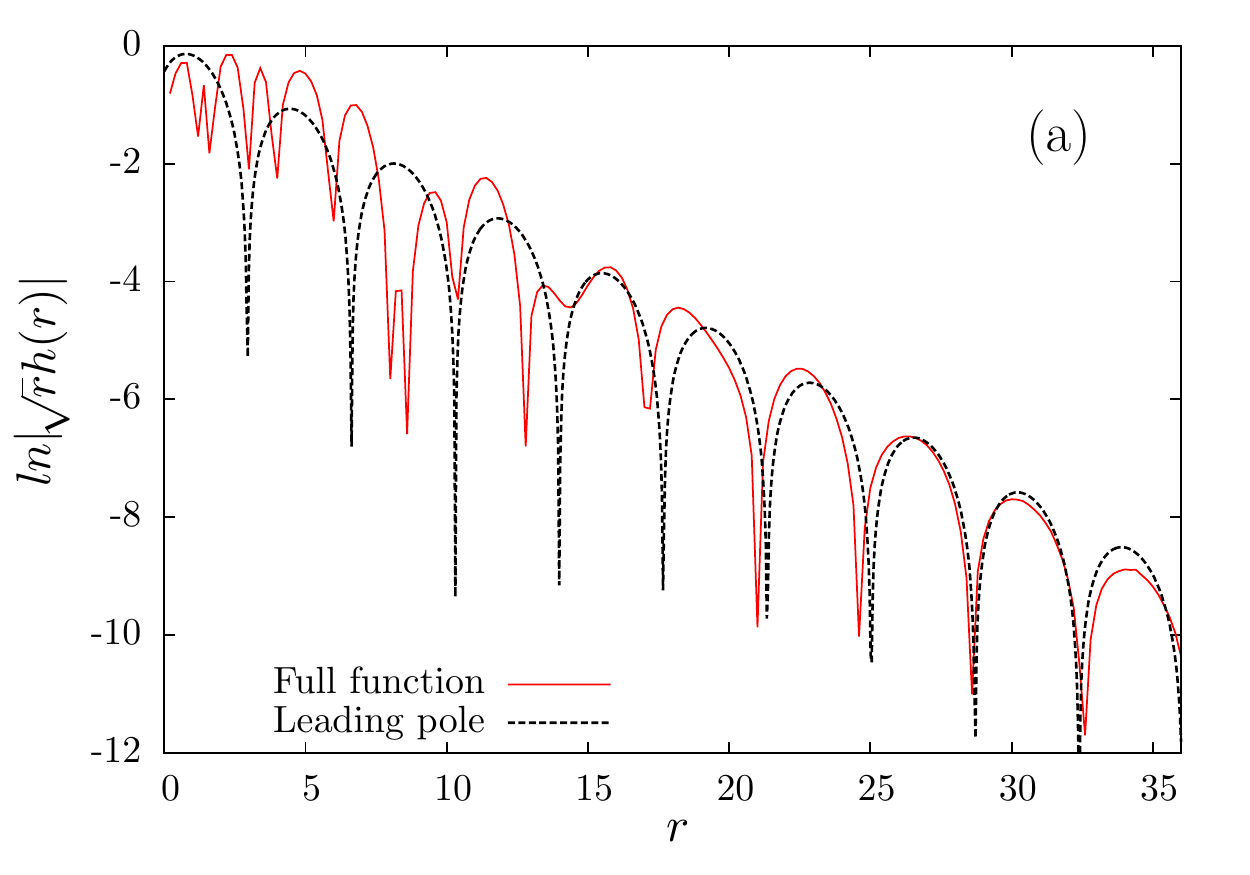}
      \includegraphics[width=0.99\columnwidth]{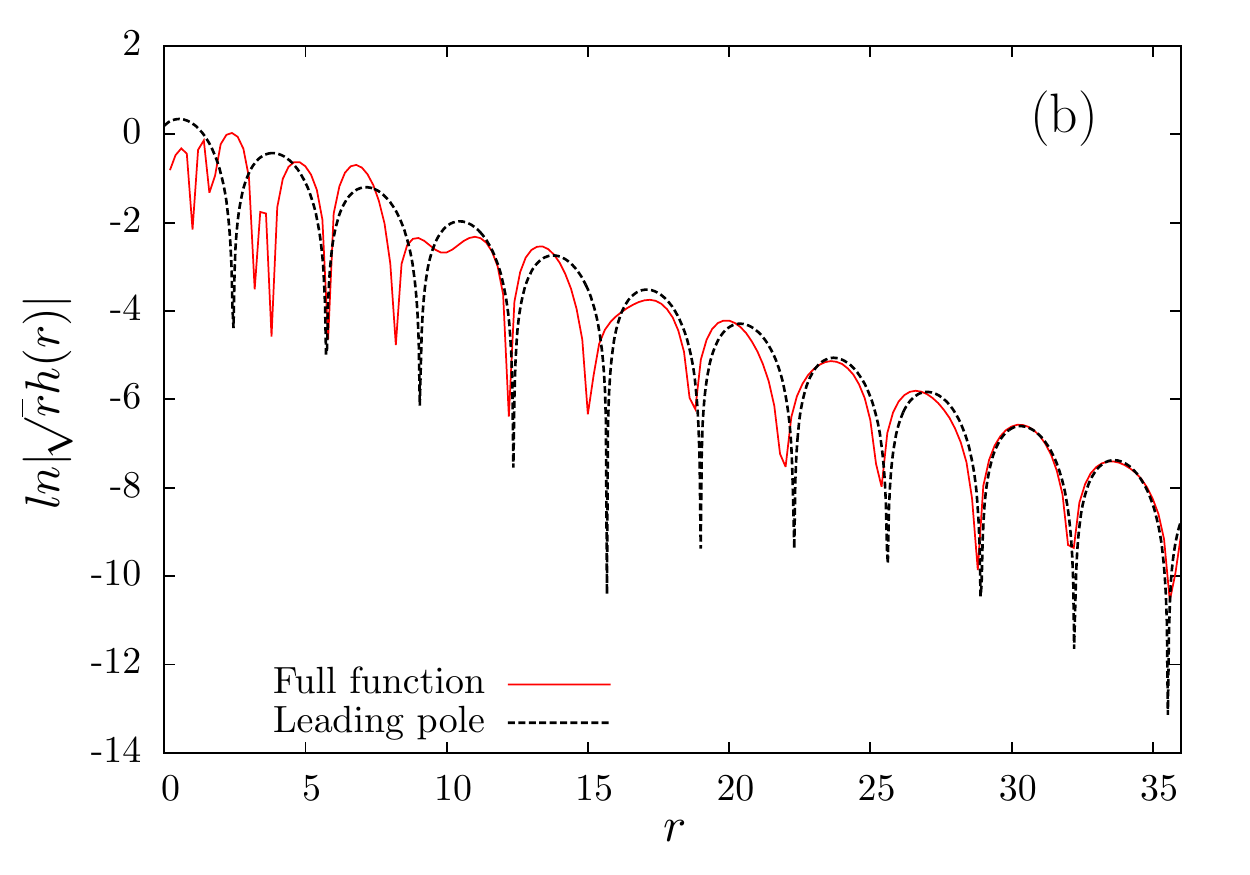}
     \caption{The asymptotic decay of $h(r)$ for $\beta\epsilon=10$ and two low density states. The solid red line denotes the results of the full RPA-DFT test particle calculations while the black dashed line corresponds to the `leading pole' approximation Eq.\,\eqref{eqn20}. {(a) With $\rho_0 =0.09$ and $C_4 =0.42$; the leading pole is an inner one with ${\alpha}_0=0.2528$ and $\alpha_1 =0.8535$.  (b) With $\rho_0 =0.08$ and $C_4 =0.44$; the leading pole is again an inner one with $\alpha_0=0.2424$ and $\alpha_1 =0.9037$.}}
      \label{fig8}
\end{figure}

\begin{figure}
\centering
   \includegraphics[width=0.99\columnwidth]{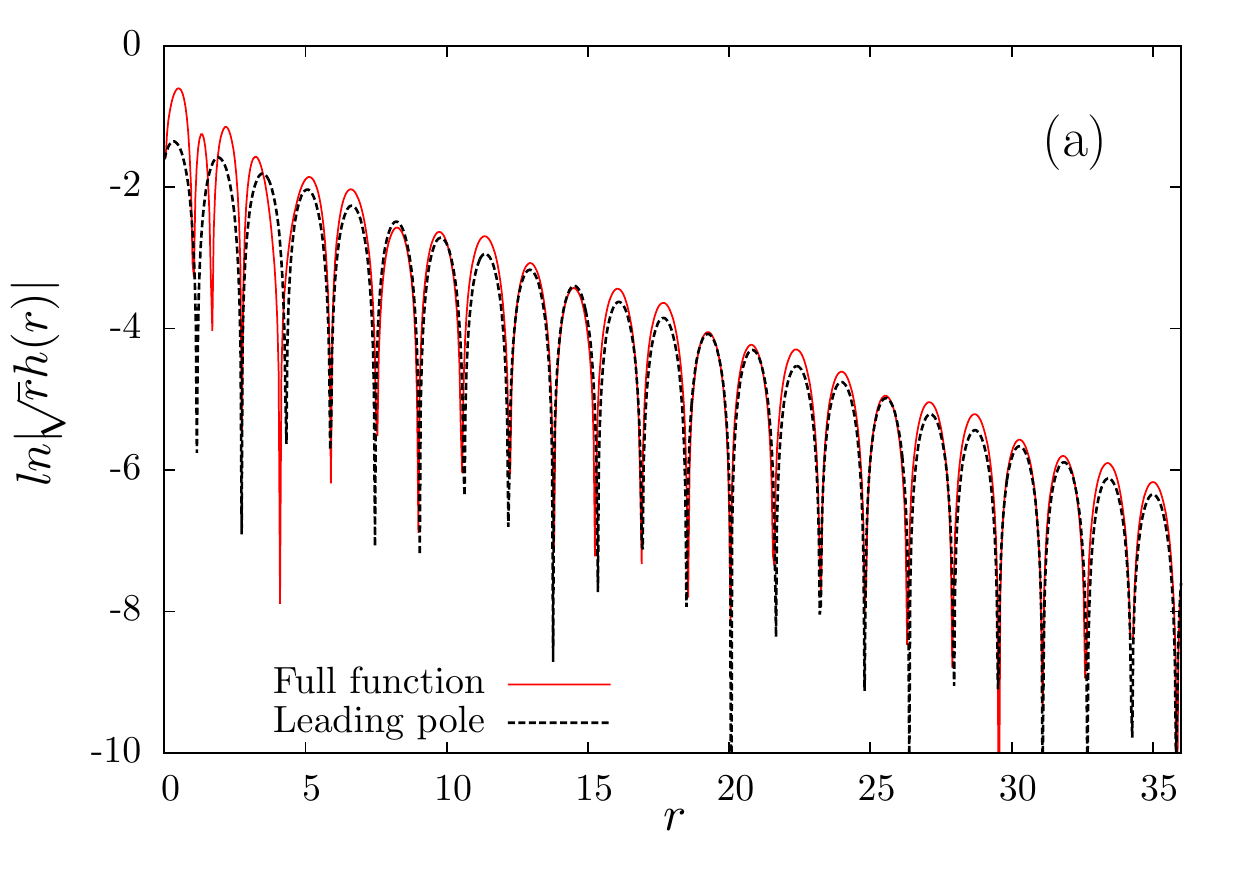}
   \includegraphics[width=0.99\columnwidth]{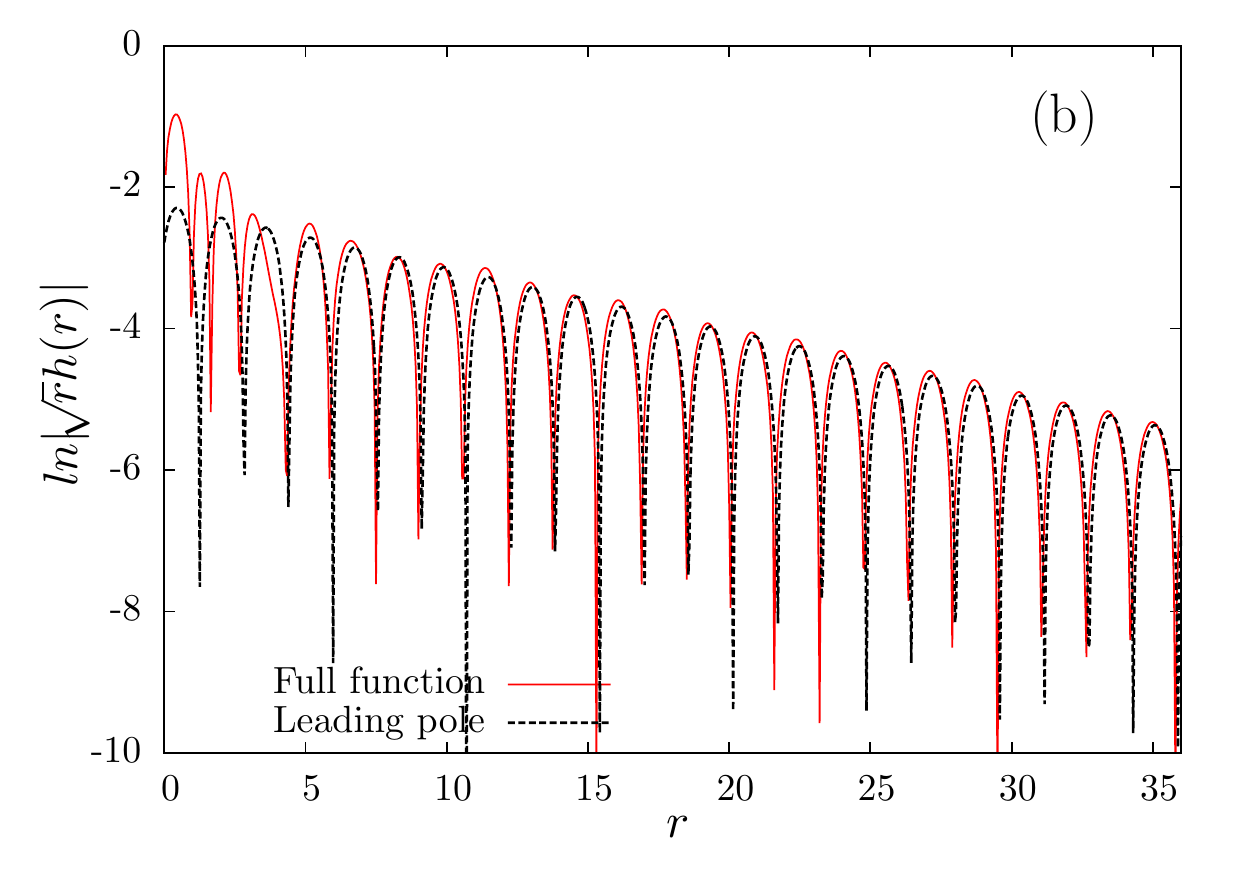}
     \caption{The asymptotic decay of $h(r)$ for $\beta\epsilon=10$, $C_4 =0.42$ and two high density states. The solid red line denotes the results of the full RPA-DFT test particle calculations while the black dashed line corresponds to the `leading pole' approximation Eq.\,\eqref{eqn20}. {(a) with $\rho_0 =0.5$; the leading pole is an outer one with $\alpha_0=0.1440$ and $\alpha_1 =1.993$. (b) with $\rho_0 =1.0$; the leading pole is again an outer one with $\alpha_0=0.0888$ and $\alpha_1 =1.996$.}}
      \label{fig9}
\end{figure}

Figs.\,\ref{fig8} and \ref{fig9} demonstrate clearly the efficacy of the asymptotic analysis. These figures plot $\ln\left[ \sqrt{r}\,h(r) \right]$ versus $r$, comparing the `full function', i.e., the results of the RPA-DFT test particle calculations, with a single `leading pole' approximation given by the first term in Eq.\,\eqref{eqn20}. Fig.\,\ref{fig8} is for two low density state points corresponding to the region in Fig.\,\ref{fig6} where the inner poles have the smallest imaginary part and dictate the asymptotic decay. For $r \gtrsim20$ the `leading pole' approximation captures accurately both the wavelength and the decay length of the oscillations; note that we match the amplitude and phase in Eq.\,\eqref{eqn20} to the numerical results. The wavelength is $\approx 2\pi$ in both cases. For the two high density states in Fig.\ 9 the outer poles dictate the decay and the `leading pole' approximation is very accurate for $r\gtrsim5$. In both cases the wavelength is $\approx0.52\times2\pi$. 

Structural crossover is not manifest in the behaviour of the two principal peaks in the structure factor $S_{RPA}(k) = \left( 1+\rho_0 \beta \hat{v}(k) \right)^{-1}$. For $C_4 ={C_{4c}}$ (see Figs.\ 2 and \ref{fig6}) the two minima in $\hat{v}(k)$ are equal and therefore the principal peaks have equal height. If $C_4 < {C_{4c}}$ the peak at the larger wavenumber is higher whereas for $C_4 > {C_{4c}}$ the peak at smaller $k$ is higher. This threshold value of $C_4 ={C_{4c}}$ defines a horizontal line in Fig.\,\ref{fig6}. Except at high densities this is well-removed from the structural crossover line. The latter terminates at the intersection of two (red) lines where the fluid first becomes linearly unstable. We return to this important feature in the next section.

{Note that generally the values of $\alpha_1$ are not exactly at the minima of $\hat{v}(k)$, since $\alpha_1$ is the real part of the complex pole solutions to Eq.~\eqref{eqn18}. In practice, $\alpha_1$ lies close to the minima of $\hat{v}(k)$ for lowest poles, with small $\alpha_0$ values. This can also be seen from combining equations \eqref{eqn3} and \eqref{eqn4} to give
\begin{equation}
\hat{h}_{RPA}(k) = \frac{-\beta \hat{v}(k)}{1+\rho_0\beta \hat{v}(k)}
\end{equation}
i.e.\ the maxima of $\hat{h}_{RPA}(k)$ are close to the minima of $\hat{v}(k)$ and so the latter are close to the least damped modes in $h(r)$.}

\section{Stability of the uniform fluid and quasicrystal formation}\label{sec4}

The analysis described in Sec.~3 focuses on the stable fluid region of the phase diagram.  Here we discuss what occurs as the density and/or the parameter $C_4$ increase in the phase diagram of Fig.\,\ref{fig6}.  We note from the first row of Fig.\,\ref{fig6} and from Fig.\,\ref{fig7} that the imaginary part $\alpha_0$ of the leading (outer) pole decreases with increasing density at fixed $C_4 = 0.42$ and that $\alpha_0\rightarrow 0^+$ for $\rho_0\approx2.5$. In this limit the pole determined by Eq.\,\eqref{eqn18} is real and the leading decay of $h(r)$ becomes undamped oscillatory, with wavelength $2\pi/\alpha_1\approx \pi$, signalling an instability of the uniform fluid. Similarly, for the second row of Fig.\,\ref{fig6} one observes $\alpha_0$ for the leading (inner) pole decreasing towards zero at fixed $\rho_0=0.744$ as $C_4$ increases indicating a possible instability. This pole analysis for determining an instability is equivalent to the venerable Kirkwood-Monroe \cite{kirkwood1941statistical} approach for tackling freezing; see also Ref.\ \cite{henderson1994instability}. It is based, of course, on static (equilibrium) considerations.

 \begin{figure}[t]
\centering{
   \includegraphics[width=0.99\columnwidth]{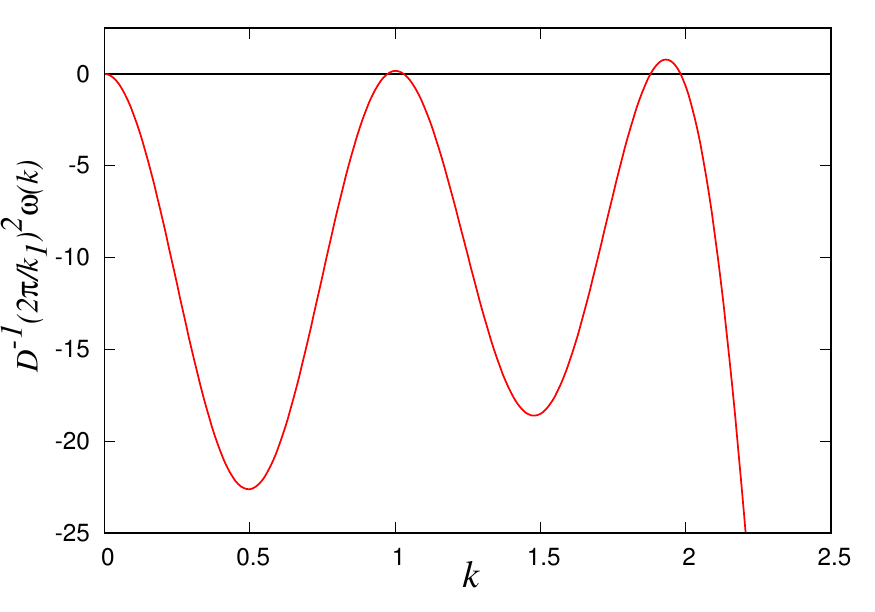}}
     \caption{The dispersion relation for the BEL potential at the state point $\beta\epsilon=10$, $C_4=0.44$, $\rho_0= 1.3$. As the maxima at $k\approx1.0$ and at $k\approx1.93$ are positive, density modulations with both wavenumbers will grow; this state is linearly unstable.}
      \label{fig10}
\end{figure}
On general grounds we can expect dynamical instabilities to occur at real wavenumbers $k>0$ as the density of the fluid, or $C_4$, is increased. There are several approaches, but we follow closely that adopted recently by Archer et al.\ \cite{archer2013quasicrystalline, archer2015soft} who considered a polymeric system with a soft core plus a corona (or shoulder) architecture, modelled as a sum of two repulsive GEM-8 potentials of different ranges and strengths, and investigated the time evolution of this model using DDFT (Dynamical DFT) -- an accurate approximation for soft particles undergoing Brownian (overdamped) stochastic dynamics. In DDFT the time ($t$) evolution of the one-body density $\rho(\rr,t)$ is given by the deterministic equation 
\begin{equation}
\frac{\partial \rho(\rr,t)}{\partial t} = \Gamma \nabla\cdot\left[ \rho(\rr,t) \nabla \frac{\delta \Omega[\rho(\rr,t)]}{\delta \rho(\rr,t)} \right]\label{eqn21}
\end{equation}
where $\Gamma$ is a mobility coefficient and $\Omega[\rho]$ is the same grand potential functional as in equilibrium DFT \cite{marconi1999dynamic, archer2004dynamical}. We are concerned with the growth of density fluctuations when a uniform fluid of density $\rho_0$ is weakly perturbed, i.e.\ we consider a small amplitude perturbation of the density at early times. Making a functional Taylor expansion of the excess free energy functional $\mathcal{F}_{ex}[\rho]$ together with appropriate linearization one obtains the following result \cite{archer2013quasicrystalline, archer2015soft, archer2004dynamical, archer2014solidification, archer2012solidification} for the Fourier decomposition of the density perturbation:
\begin{equation}
\tilde{\rho}(\rr,t)\equiv \rho(\rr,t)-\rho_0 = \sum_{\kk}\hat{\rho}(\kk,t=0) e^{i\kk\cdot\rr+\omega t},\label{eqn22}
\end{equation}
with the dispersion relation
\begin{equation}
\omega(k) = -Dk^2[1-\rho_0\hat{c}(k)]\label{eqn23}
\end{equation}
where the diffusion coefficient is $D = k_B T \Gamma$ and $k=|\kk|$. Note that an equivalent result was obtained in an early DFT treatment of spinodal decomposition \cite{evans1979nature, evans1979spinodal}. That the uniform fluid direct correlation function enters Eq.\,\eqref{eqn23} is a direct consequence of the functional Taylor expansion of the free energy functional about the uniform density $\rho_0$. For a \textit{stable} uniform fluid the OZ Eq.\,\eqref{eqn4} implies $[1-\rho_0\hat{c}(k)]^{-1} = S(k)$, the liquid structure factor. The usual stability criterion that $S(k)$ must be positive for all real wavenumbers $k$ is therefore equivalent to requiring $\omega(k)\leq 0$ for all $k$. This result implies all Fourier modes in Eq.\,\eqref{eqn22} must decay with increasing time: such a state is linearly stable. Recalling the definition \eqref{eqn6} of the direct correlation function as a second functional derivative, it is clear there will be bulk state points where $[1-\rho_0\hat{c}(k)]<0$  for certain $k$ \cite{archer2004dynamical}. Correspondingly, $\omega(k)>0$ and the mode grows with time: the state is linearly unstable. Such states occur inside the parameter regime where a crystal is the equilibrium phase. The onset of linear instability, or the marginal stability threshold, is given by the locus in the phase diagram where the maximum growth rate is zero. This is defined by the two conditions
\begin{equation}
\frac{d\omega(k)}{dk}\Big|_{k=k_c} \, = 0 \,\,\,\,\textrm{and}\,\,\,\,\omega(k=k_c) = 0\label{eqn24}
\end{equation}
where $k_c$ is the wavenumber of the marginally unstable mode. 

We now make the connection between the pole analysis and the present linear stability investigation. The latter dictates, from Eq.\,\eqref{eqn24}, that a linear instability occurs when $[1-\rho_0\hat{c}(k_c)]=0$, with $k_c$ real. But this is just the criterion for a pole (see Eq.\,\eqref{eqn18} and below) with vanishing imaginary part, and $k_c=\alpha_1$, for the purely real pole. For example, the decrease of $\alpha_0$ towards zero that we see in Fig.\,\ref{fig7} corresponds to the approach to the onset of linear instability as determined by the conventional dispersion relation. We can now consider the genesis of the (red) instability lines in Fig.\,\ref{fig6}.

Within the simple RPA, $\hat{c}(k) = -\beta \hat{v}(k)$ is independent of density and thus the Fourier transform of the pair potential determines directly the form of the dispersion relation. It follows from Eq.\,\eqref{eqn24} that for a given $C_4$, the wavenumber $k_c$ corresponds precisely to a minimum in $\hat{v}(k)$. In Fig.\,\ref{fig6} the lower red line is the high-density branch of the onset of instability and is associated with the second minimum of $\hat{v}(k)$, occurring at the larger wavenumber. On this line $k_c$ corresponds to the second maximum of $\omega(k)$, near $k_2$, reaching zero while the upper line is the low-density branch where the first maximum of $\omega(k)$, near $k_1$, reaches zero. For $C_4 ={C_{4c}}$ the two minima in $\hat{v}(k)$ are equal and the two branches meet at the density $\rho_0\approx1.25$. At this point the two maxima in the dispersion relation are both zero and modes with the values $k_1 =1$ and $k_2 =1.932$ grow initially at the same rate. The scenario presented here is close to that in Fig.\ 1 of Refs.\ \cite{archer2013quasicrystalline, archer2015soft}. Our parameter $C_4$ plays the role of their parameter $a$ which determines the strength of the corona repulsion. As in Refs.\ \cite{archer2013quasicrystalline, archer2015soft}, we find the two branches cross. However, for clarity, the smooth extensions of the branches beyond the crossing point are not shown in our Fig.\,\ref{fig6}.

 \begin{figure*}
\centering
   \includegraphics[width=0.8\columnwidth]{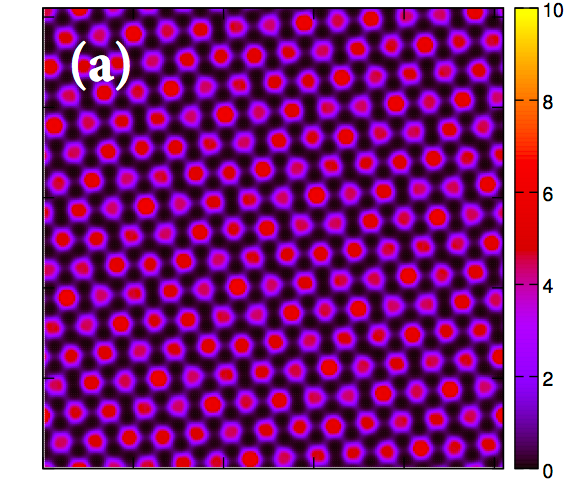}
      \includegraphics[width=0.8\columnwidth]{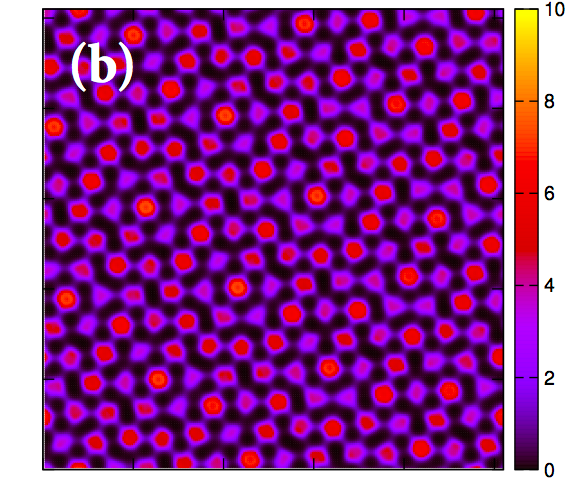}
      
      \includegraphics[width=0.8\columnwidth]{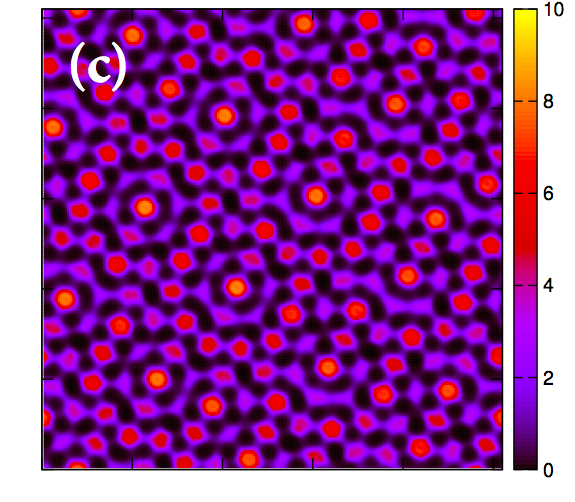}
      \includegraphics[width=0.8\columnwidth]{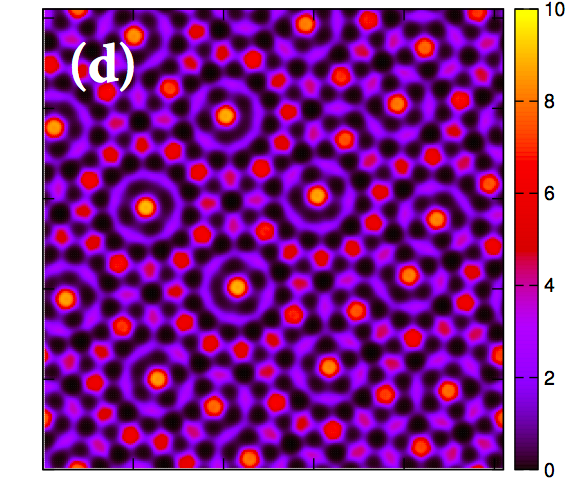}   
     \caption{DDFT results for the time evolution of the density profile in the $(x, y)$ plane following a quench to the uniform state with $\beta\epsilon=10$, $C_4=0.44$, $\rho_0= 1.3$. The corresponding dispersion relation is displayed in Fig.\,\ref{fig10}. The profiles are shown {(a)} for $t^*=16$ (top left), {(b)} $t^*=40$ (top right), {(c)} $t^*=80$ (bottom left) and {(d)} $t^*=200$ (bottom right), where $t^*=t/\tau_B$, with Brownian time $\tau_B = (2\pi/k_1)^2/D$. In the earlier stages the density modulations display a shorter length scale (see text) while at the latest time quasicrystalline structure is clearly present. }
      \label{fig11}
\end{figure*}

In Fig.\,\ref{fig10} we plot the dispersion relation at the state point $\beta\epsilon=10$, $C_4=0.44$ and density $\rho_0= 1.3$. This density is slightly larger than that where the two instability branches meet. At this state point the two almost-equal minima in $\hat{v}(k)$ at $k_1 =1$ and $k_2 =1.932$ yield two maxima in $\omega(k)$ at wavenumbers that are very close to these values. Since $\omega(k)$ is positive at both maxima the fluid at this state point is linearly unstable with respect to density fluctuations with both wavenumbers, albeit weakly at the smaller wavenumber. `Quenching' the uniform fluid to this state point will lead to non-uniform structures as the system evolves in time. In Fig.\,\ref{fig11} we show density profiles computed from DDFT, Eq.\,\eqref{eqn21}, following such a quench. In implementing the DDFT time evolution we add, at $t=0$, a small amplitude random fluctuating variable $\xi(\rr)$ to the uniform density $\rho_0$ at each point in space, i.e., $\rho(\rr,t=0) = \rho_0 + \xi(\rr)$;  see \cite{archer2013quasicrystalline, archer2015soft}. As the maximum in $\omega(k)$ is larger at $k\approx1.932$ than at $k\approx1$ we see that the density modulations with the larger wavenumber grow faster initially than those for the smaller value. At the early time $t^* =16$ ({Fig.\,\ref{fig11}(a)}) we find modulations with a short length scale, i.e., we observe a hexagonal crystal with a short periodicity. At subsequent times non-linear evolution involves both length scales. The final structure ({Fig.\,\ref{fig11}(d)}) at $t^*=200$ is the local equilibrium state of our system. This displays domains of dodecagonal quasicrystal ordering. The sequence of structures we observe is somewhat similar to that found in \cite{archer2013quasicrystalline, archer2015soft} for the double GEM-8 model potential at  state points for which the dispersion relation is similar (see bottom panel of Fig.\,12 in \cite{archer2015soft}) to our Fig.\,\ref{fig10}. For the double GEM-8 model, the full phase diagram was determined  \cite{archer2013quasicrystalline, archer2015soft}. One finds that for states where the dispersion relation is of this form the system first forms a crystal with a short length scale, since that is the most unstable fastest growing mode. However, in the double GEM-8 system such a crystal does not correspond to the equilibrium state, which is in fact a longer length scale crystal, whereas for the present BEL model the QC is the minimum free energy state for some parameter vlaues.

Suppose now we follow state points on the structural crossover line in Fig.\,\ref{fig6}, starting deep in the equilibrium fluid phase, and move towards higher densities and higher $C_4$. The inner and outer poles have real parts that approach $k_1$ and $k_2$ while the common imaginary part $\alpha_0$ decreases towards zero. On the other hand, if we consider the two lines of linear instability determined by Eq.\,\eqref{eqn24} we find that at the point of their intersection (at $C_4 ={C_{4c}}$ and $\rho_0\approx1.25$) the minima in $\hat{v}(k)$ are equal. It follows that within our RPA treatment,
\begin{equation}
1-\rho_0\hat{c}(k_1) = 1-\rho_0 \hat{c}(k_2) =0\label{eqn25}
\end{equation}
for the (real) wavenumbers $k_1$ and $k_2$. Thus, at some special state point there are inner and outer poles, both with vanishing imaginary part, whose real parts are $k_1$ and $k_2$. It is striking that the region where quasicrystals form lies in the neighbourhood of this point.

We note that in the stable region of the phase diagram in Fig.\,\ref{fig6} it can be useful to think in terms of the structure factor $S_{RPA}(k) = [1+\rho_0\beta\hat{v}(k)]^{-1}$. For example, if we follow the horizontal line at $C_4 = {C_{4c}}$ increasing the density, this function is positive and finite for all (real) $k$, with equal principal peaks at $k_1$ and $k_2$, until reaching the point of intersection near $\rho_0=1.25$. Then $S_{RPA}(k)$ diverges at both wavenumbers, consistent with Eq.\,\eqref{eqn25}. {Note also that a line in the phase diagram along which the static structure factor diverges for $k\neq0$ is sometimes referred to as the lambda line, particularly in the context of ionic liquids \cite{stell1995criticality, ciach2003effect, kirkwood1936jg, archer2004soft, archer2007model}.}

\section{Discussion and Conclusions}\label{sec5}

We have determined the liquid state structure of a model fluid composed of particles interacting via the BEL pair potential \eqref{eqn1}. At sufficiently high densities and for particular values of the pair potential parameters $\{C_n,\sigma\}$ it is known that the system solidifies to form a QC \cite{barkan2014controlled}. Here, we find that the propensity towards QC formation is manifest in the liquid state structure. In particular, the decay $r\to\infty$ of the radial distribution function $g(r)$ contains two exponentially damped oscillatory contributions with quite different wavelengths but similar decay lengths. These are associated with distinct peaks, at wavenumbers $k_1$ and $k_2$, in the static structure factor $S(k)$, where $k_2/k_1\approx1.932$. The double peaked form in $S(k)$ with the same ratio was also observed for a very different model system, namely a double GEM-8 model \cite{archer2013quasicrystalline, archer2015soft} designed to mimic a polymeric system with a soft core plus corona architecture. We believe that these features in $S(k)$ and $g(r)$ should be generic to QC forming systems. Identifying these in the liquid state will provide useful sign-posts to finding other systems that solidify to form QC. Our analysis shows that there is a cross-over line in the liquid-state portion of the phase diagram at which the asymptotic decay of $g(r)$ changes from damped oscillatory decay with wavelength $\approx2\pi/k_1$ to damped oscillatory with different wavelength $\approx2\pi/k_2$. Following the locus of this line towards higher density states leads directly to the portion of the phase diagram where QC occur; see Fig.\,\ref{fig6}. {In previous studies seeking to find model systems that form QCs \cite{barkan2011stability, archer2013quasicrystalline, barkan2014controlled, archer2015soft, subramanian2016three, subramanian2017spatially} the strategy used was to identify state points where the dispersion relation has the required double peaked structure. This is akin to identifying a double peaked shape in $S(k)$. The new insight from the present study is that a search strategy based on examining the real-space liquid state correlations would also be at least as effective}.

In determining $g(r)$ for the BEL model we have used both HNC theory and the RPA-DFT test particle route. The excellent agreement between the two (see Fig\,\ref{fig34}) indicates that the simpler RPA-DFT is rather accurate. This result was not obvious, given the complex form of the pair potential, which contains multiple length and energy scales (see Fig.\,\ref{fig1}). Although the pole analysis to determine the asymptotic decay form of $g(r)$ for $r\to\infty$ was performed solely for the RPA, we do not expect the results we have obtained to be significantly different from those one would obtain with HNC or any other reliable integral equation theory or simulation.

The pole analysis used here to determine asymptotic decay of $g(r)$ is a generalisation to 2D of an approach that has previously been used successfully for 3D systems. Here, we have shown that the general form of the asymptotic decay $r\to\infty$ of $g(r)$ for 2D fluids, with short-ranged interparticle potentials away from the critical point, is either of the form in Eq.\,\eqref{eqn19} or that in Eq.\,\eqref{eqn20}. Whilst these results could have been guessed, based on our knowledge of the well known results in 3D, Eqs.\,\eqref{eqn8} and \eqref{eqn9}, the mathematical derivation in 2D is somewhat different from in 3D. In particular, the steps in Eqs.\,\eqref{eqn12}--\eqref{eqn16} are particular to 2D. Therefore, the present work provides a valuable contribution to the study of 2D fluids in general.

As described in the introduction, structural crossover in the asymptotic decay of pair correlation functions is not unexpected in binary mixtures when there is a sufficiently large difference in the sizes of the two species of particles \cite{archer2001binary, grodon2004decay, grodon2005homogeneous}. However, the presence of two different length scales often needs to be engineered in one-component mixtures. Therefore, in this regard the present BEL model is unusual and suggests why one component systems that form QC are not common. We anticipate that binary mixtures having a crossover in the asymptotic decay of the three partial radial distribution functions $g_{ij}(r)$ from oscillatory decay with wavenumber $k_1$ to oscillatory decay with wavenumber $k_2$, with $k_2 /k_1 = 2\cos(\pi/n)$ and with $n =12$, will in 2D be candidates for forming dodecagonal QC. Other values of $n$ will also be interesting. We believe that binary colloidal mixtures, where the diameters of the colloids can be finely tuned \cite{baumgartl2007experimental, statt2016direct}, will be the most likely candidates for investigation.

Finally, we highlight the connection, made explicit here, between the pole analysis for the static equilibrium fluid structure that is based on finding zeros of the quantity $[1-\rho_0 \hat{c}(k)]$ in the complex-$k$ plane and the study of the non-equilibrium growth or decay of density modulations, which is determined by the dispersion relation $\omega(k)$ in Eq.\,\eqref{eqn23}. $\omega(k)$ is proportional to exactly the same quantity, but evaluated for real values of $k$. The connection is due to the fact that both approaches are based on a linear response treatment. Both emphasise the importance of the quantity $\hat{c}(k)$, or its Fourier transform to real space, $c(r)$, defined in Eq.\,\eqref{eqn6}.

\section*{Acknowledgements}
It is a pleasure to dedicate this paper to two truly inspiring figures in liquid state science on their $70^{th}$, Daan Frenkel, and $90^{th}$, Ben Widom, birthdays. Daan led us to ponder the novelty, and even the existence, of a stable liquid state. Ben taught us why correlation functions are important and fascinating! M.C.W. was supported by an EPSRC studentship, P.S. was supported in part by a L'Or{\'e}al UK and Ireland Fellowship for Women in Science, A.J.A. was supported by EPSRC Grant No. EP/P015689/1 and R.E. was supported by Leverhulme Trust Grant No. EM-2016-031.
 

%

\end{document}